\documentclass[journal]{IEEEtran}
\usepackage{amssymb}
\usepackage{stfloats}
\usepackage{graphicx}
\usepackage{multirow}
\usepackage{subfigure}
\usepackage{tabularx}
\usepackage{booktabs}
\usepackage{amsmath}
\usepackage{amsthm}
\usepackage{epsfig,epsf,color,balance,cite}
\usepackage{indentfirst}
\usepackage{mathrsfs}
\usepackage{dsfont}
\UseRawInputEncoding
\usepackage{easyReview}
\usepackage[justification=centering]{caption}
\usepackage[font=small,labelfont=md,labelsep=period]{caption}

\newtheorem{remark}{Remark}
\usepackage{algorithm, algorithmicx, algpseudocode}
\usepackage{enumerate}
\usepackage{makecell}
\usepackage{url}
\newcounter{MYtempeqncnt}

\usepackage[colorlinks,bookmarksopen,bookmarksnumbered,citecolor=red, linkcolor=red, urlcolor=red]{hyperref}
\usepackage{graphicx}
\begin{document}
	\title{Joint Visibility Region and Channel Estimation
		for Extremely Large-Scale MIMO Systems}
	\author{
		\IEEEauthorblockN{ 
			Anzheng Tang\IEEEauthorrefmark{1},~
			Jun-Bo Wang\IEEEauthorrefmark{1},
			Yijin Pan\IEEEauthorrefmark{1},
			Wence Zhang\IEEEauthorrefmark{2},
			Xiaodan Zhang\IEEEauthorrefmark{3},
			Yijian Chen\IEEEauthorrefmark{4},\\
			Hongkang Yu\IEEEauthorrefmark{4},
			and Rodrigo C. de Lamare\IEEEauthorrefmark{5}
			\vspace{-1em}
		}
		\thanks{Anzheng Tang, Jun-bo Wang and Yijin Pan are with the National Mobile Communications Research Laboratory, Southeast University, Nanjing 210096, China. (email: \{anzhengt, jbwang, and panyj\}@seu.edu.cn)
		}
		\thanks{Wence Zhang is with the School of Computer Science and Communication Engineering and the Jiangsu Key Laboratory of Security Technology for Industrial Cyberspace, Jiangsu University, Zhenjiang 212013, China. (e-mail: wencezhang@ujs.edu.cn)}
		\thanks{Xiaodan Zhang is with the School of Management, Shenzhen Institute of Information Technology, Shenzhen 518172, China. (e-mail: Dannyzxd@163.com)} 
		\thanks{Yijian Chen and Hongkang Yu are with the Wireless Product Research and Development Institute, ZTE Corporation, Shenzhen 518057, China. (e-mail:\{yu.hongkang, chen.yijian\}@zte.com.cn)}
		\thanks{Rodrigo C. de Lamare is with the Centre for Telecommunications Studies, Pontifical Catholic University of Rio de Janeiro, Rio de Janeiro 22451-900, Brazil, and also with the Department of Electronic Engineering, University of York, York YO10 5DD, U.K. (e-mail: delamare@puc-rio.br).}
	}
	\maketitle
	\vspace{-3em}
\begin{abstract}
	In this work, we investigate the joint visibility region (VR) detection and channel estimation (CE) problem for extremely large-scale multiple-input-multiple-output (XL-MIMO) systems considering both the spherical wavefront effect and spatial non-stationary (SnS) property. {Unlike existing SnS CE methods that rely on the statistical characteristics of channels in the spatial or delay domain, we propose an approach that simultaneously exploits the antenna-domain spatial correlation and the wavenumber-domain sparsity of SnS channels.} To this end, we introduce a two-stage VR detection and CE scheme. In the first stage, the belief regarding the visibility of antennas is obtained through a VR detection-oriented message passing (VRDO-MP) scheme, which fully exploits the spatial correlation among adjacent antenna elements. In the second stage, leveraging the VR information and wavenumber-domain sparsity, we accurately estimate the SnS channel employing the belief-based orthogonal matching pursuit (BB-OMP) method. Simulations show that the proposed algorithms lead to a significant enhancement in VR detection and CE accuracy as compared to existing methods, especially in low signal-to-noise ratio (SNR) scenarios.
\end{abstract}   
	
\IEEEpeerreviewmaketitle
	
	\section{Introduction}
	\label{section1}
    Extremely large-scale multiple-input-multiple-output (XL-MIMO) systems operating in millimeter wave (mmWave) or sub-terahertz (sub-THz) bands has been widely regarded as a promising technique to meet the explosive data demand in beyond 5G (B5G) and 6G networks \cite{Zeng, ZTE1,ZTE2}. In comparison to traditional massive MIMO systems, XL-MIMO systems increase the number of antennas by an order of magnitude, which significantly enhances system performance in terms of capacity, spectral efficiency, transmission delays and reliability \cite{MultipleAntennaTechnologies, XL_MIMO3, XL_MIMO_R1, XL_MIMO_R2}. Therefore, the concepts of XL-MIMO have attracted substantial attention from academia and industry. 
	\begin{figure}
		\centering
		\includegraphics[width=0.45\textwidth]{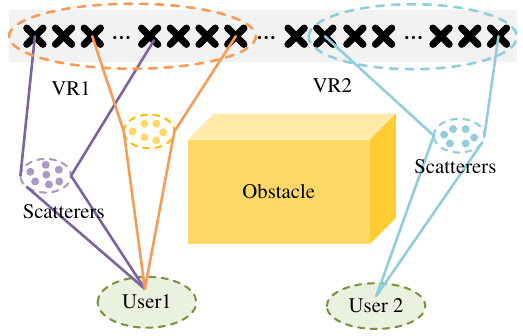}
		\caption{Illustration of spatial non-stationarity in XL-MIMO systems}
		\label{VRs}
	\end{figure}
	
	In XL-MIMO systems, the large-aperture array results in a more significant Rayleigh distance, causing the transmission to occur in the near-field region \cite{LoS_Angular, HolographicMIMORayleigh,NF2}. {For example, with an antenna diameter of $L=0.1$ m at a carrier frequency of $5$ GHz, any receiver located at a distance greater than $0.33$ m is considered to be in the far-field region. However, in terms of mmWave bands, for an antenna diameter of $L=0.5$~m at a carrier frequency of $28$ GHz, any user closer than 47 m from the antenna resides in the near-field region. It can be seen that the near-field effect is more significant in mmWave bands, particularly when combined with antenna arrays of relatively large size compared to traditional microwave bands.}
	In this case, the planar wavefront assumption used in far-field communications becomes invalid, as the wavefront curvature over the array is no longer negligible. Additionally, due to obstacles and incomplete scattering, spatial non-stationary (SnS) property may appear along the array \cite{SnS1}. Specifically, antennas at different spatial positions may observe distinct channel multipath characteristics, as shown in Fig. \ref{VRs}. To quantitatively characterize the SnS property, the visibility region (VR) based channel modeling method has been adopted  in XL-MIMO systems \cite{HanYu, SnS1, SnS2, SnS3}.
	 
The presence of spherical wavefront effects and the SnS property in XL-MIMO systems opens up new possibilities for spatial multiplexing and random access (RA), respectively.
Due to the non-negligible wavefront curvature over the array, even in a line-of-sight (LoS) scenario, multiple spatial frequencies are associated with the LoS path between transceiver arrays. Consequently, XL-MIMO channels exhibit a high-rank property, enabling the transmission of multiple data streams \cite{LoS_Angular, LoSMIMO_HighRank, DoF1, SpatialBandwidth2, LOSMIMO2, R_ULAs}. In our previous work \cite{LoS_Angular}, we also evaluated the high-rank characteristics in the wavenumber domain. To adapt the channel ranks in the case of different signal-to-noise ratio (SNR) values and enhance system performance, \cite{R_ULAs} proposed reconfigurable uniform linear array (ULA) schemes.
Additionally, the spherical wavefront introduces extra resolution in the distance domain. Capitalizing on these spatial resources in the distance domain, \cite{LDMA} proposed a novel location division multiple access (LDMA) scheme to simultaneously serve users at different locations. Despite spatial interference between beam focusing vectors at different positions, appropriate partitioning of spatial resources allows for minimizing the correlation between beam vectors corresponding to different sampling points. Compared to classical far-field spatial division multiple access (SDMA) schemes, the LDMA scheme exploits the focusing of near-field beams rather than the steering property of far-field beams. This offers a  method to efficiently exploit spatial resources, enabling support of more users for simultaneous access.
Furthermore, in XL-MIMO systems the heightened user density, coupled with limited pilots, poses challenges in random access and scheduling protocols. Fortunately, the SnS property provides a unique advantage in enhancing connectivity performance \cite{VRsAccess}. Specifically, users with non-overlapping VRs can be scheduled within the same payload data pilot (PDP) resource. This approach effectively alleviates the pilot scarcity issue, resulting in significant improvements in the overall sum rate compared to conventional random access schemes. 	

Notably, the successful implementation of spatial multiplexing and non-overlapping VRs based random access protocol depends on accurate channel and VR information. Consequently, the joint detection of VR and channel estimation (CE) becomes a critical topic in XL-MIMO systems.
\vspace{-1em}
\subsection{Related Works}
Most existing works, such as \cite{PolarCS, NFBT1, NFBT2, NFCE1, NFCE2}, primarily focused on spherical wavefront effect and did not consider the SnS property. Specifically, \cite{PolarCS} proposed a novel CE scheme based on the polar-domain sparsity of XL-MIMO channels. Based on the polar-domain codebook, on-grid and off-grid simultaneous orthogonal matching pursuit (OMP) algorithms were proposed to accurately reconstruct the channels. Then, the polar-domain codebook was applied to near-field beam training \cite{NFBT1, NFBT2} and hybrid-field CE \cite{NFCE1, NFCE2}. Regrettably, overlooking the SnS property could result in a significant mismatch between existing CE schemes and the SnS channels, thereby leading to a degradation in estimation performance.
	
{To address the SnS channel estimation problem, \cite{Hanyu2} proposed the subarray-wise and scatterer-wise schemes to estimate the SnS channels. Nevertheless, the schemes were heuristic without exploiting the specific channel characteristics, such as sparsity in the transformation domain. 
Moreover, based on the assumption of subarray-wise VRs, \cite{VR_wise_SnS} proposed a group time block code (GTBC) based signal extraction scheme for each subarray. Subsequently, the SnS channel estimation was transformed into several spatially stationary estimation tasks. 
However, the method may have inadvertently overlooked the spatial correlation among subarrays. Additionally, the assumption of complete subarray-wise VRs in \cite{Hanyu2, VR_wise_SnS} might be considered somewhat idealistic, as practical scenarios often involve users or scatterers with visibility limited to specific array elements within each subarray.}
In the context of SnS reconfigurable intelligent surface (RIS) cascaded channels, a three-step VR detection and CE scheme was proposed in \cite{HanYu}. Specifically, the cascaded channel was first roughly estimated according to the method in \cite{MJCE}, then the VR of the user was identified by exploiting the statistical characteristic of the received power across the antenna elements. The channel estimate was then further refined by exploiting the near-field characteristics and jointly utilizing the pilots of multiple users. However, the VR detection method based on power detection is dependent on the accuracy of CE and is sensitive to the noise level. In particular, under low SNR scenarios, this approach could experience significant performance degradation.

{Considering the inherent antenna-domain or delay-domain sparsity in SnS channels, various Bayesian inference-based methods have been proposed. Exploiting delay-domain sparsity, \cite{SnS4} introduced an adaptive grouping sparse Bayesian learning (AGSBL) scheme for uplink channel estimation. However, the sparse Bayesian learning method involved matrix inverse operation, which lead to huge computational complexity with the increasing on the number of antennas in XL-MIMO systems. Utilizing antenna-domain sparsity resulting from the SnS property, \cite{Bayesian3} characterized the XL-MIMO channels with a subarray-wise Bernoulli-Gaussian distribution. Then, a bilinear message passing (MP) algorithm was proposed for joint user activity detection and CE scheme.  To simultaneously capture antenna- and delay-domain sparsity, \cite{Bayesian1} proposed a structured prior model with the hidden Markov model (HMM) to capture the characteristics of   VR and delay domain clustering.}
However, the above solutions only used the statistical distribution (mean and variance) to characterize the spatial-domain channel while ignoring the sparsity in the transformation domain, which leads to potential estimation performance degradation. Moreover, these methods are developed based on a full digital precoding architecture, making it challenging to extend them to a hybrid precoding architecture.
\vspace{-1em}
\subsection{Main Contributions}
To solve the aforementioned problems, in this paper, we develop a joint VR detection and CE approach for XL-MIMO systems with a hybrid precoding architecture. Unlike existing works that primarily harness antenna-domain characteristics, our approach seeks to fully exploit both the antenna-domain and wavenumber-domain characteristics inherent in SnS channels. The main contributions are summarized as follows.
\begin{itemize}
	\item We first investigate the characteristics of SnS XL-MIMO channels in the antenna and wavenumber domains. We further validate the effectiveness of the VR-based SnS channel model through ray-tracing simulations, while the effects of the SnS property on the wavenumber-domain characteristics are systematically examined.

	\item To improve VR detection and CE performance, we propose a novel approach that simultaneously exploits the antenna-domain spatial correlation of the VR and the wavenumber-domain sparsity. Given the distinct domains of the SnS property and wavenumber-domain sparsity, the concurrent estimation of the VR vector and wavenumber-domain channels presents challenges. To address this, we propose a two-stage scheme to transform the concurrent estimation into a sequential estimation.
		
	\item In the first stage, the VR detection is  formulated as a Bayesian inference problem. To effectively solve this problem, we introduce a VR detection-oriented MP (VRDO-MP) algorithm. In the second stage, building upon the estimated VR information and leveraging wavenumber-domain sparsity, we propose the belief-based orthogonal matching pursuit (BB-OMP) algorithm for accurate estimation of the SnS channel.
		  
	\item In comparison to existing state-of-the-art techniques, the proposed CE algorithm demonstrates superior performance in estimation accuracy and robustness, especially when dealing with varying sizes of VRs, which validates the efficacy and applicability of the proposed CE algorithm, regardless of SnS or spatially stationary scenarios.
\end{itemize}
\vspace{-1em}
\subsection{Organization and Notations}
Organization: The remainder of this paper is organized as follows. In Section \ref{section2}, we introduce the system and channel models. Then, in Section \ref{section3}, we present a two-stage VR detection and CE (TS-VRCE) scheme. Simulations are carried out in Section \ref{section5}, and the conclusions and future works are drawn in Section \ref{section6}.
	
Notations: lower-case letters, bold-face lower-case letters, and bold-face upper-case letters are used for scalars, vectors and matrices, respectively; 
The superscripts $\left(\cdot \right)^{\mathrm{T}}$ and $\left(\cdot \right)^{\mathrm{H}}$ stand for transpose and conjugate transpose, respectively;  
$\mathrm{diag}\left(x_1, x_2, \dotso, x_N\right)$ denotes an $N \times N$ diagonal matrix with $\{x_1, x_2, \dotso, x_N\}$ being its diagonal elements;
$\mathbf{I}_{N}$ denotes an $N \times N$ identity matrix; 
$\mathbb{C}^{M \times N}$ denotes an $M \times N$ complex matrix. 
In addition, a random vector $x \in \mathbb{C}$ drawn from the complex Gaussian distribution with mean $x_0$ and variance $v$ is characterized by the probability density function $\mathcal{CN}(x; x_0,v) = {\exp\left\lbrace-{\left|x-x_0\right|^2}/{v} \right\rbrace}/{\pi v}$.

\section{System and Channel Models}
\label{section2}
\begin{figure}
	\centering
	\includegraphics[width=0.45\textwidth]{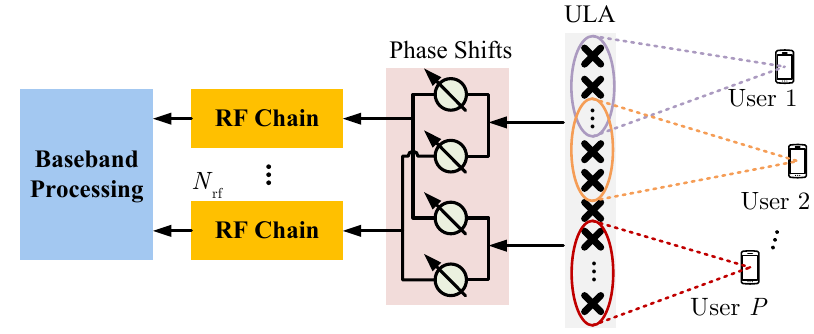}
	\caption{XL-MIMO system with hybrid precoding.}
	\label{model}
\end{figure}
Consider a time-division-duplexing (TDD)-based narrow-band XL-MIMO communication system operating in the mmWave bands\footnote{{In this manuscript, we only take mmWave bands as an example for ease of illustration. It is noted that the analysis method provided in this manuscript can be applied to THz bands in a similar way.}}, as illustrated in Fig. \ref{model}. The base station (BS) employs a hybrid precoding architecture, where an $N$-antenna uniform linear array (ULA) connects to $N_{\mathrm{rf}}$ radio frequency (RF) chains through phase shifters to serve $P$ single-antenna users. The antenna spacing is denoted by $d = \lambda/2$, where $\lambda$ represents the wavelength.

\subsection{XL-MIMO Channel Model}	
Due to the limited diffraction and severe absorption loss, the mmWave environment exhibits significantly weaker multipath propagation effects\cite{mmWaveLoS}. Consequently, our focus is primarily on the LoS path. Moreover, the combination of extremely large-scale antenna arrays and mmWave bands causes the distance between BS and users might be smaller than the Rayleigh distance. Thus, the channel response between the $n$-th antenna element and the $p$-th user with non-uniform spherical wavefront is given by \cite{Non_uniform}
\begin{equation}
	h_{p,n}=  \frac{\lambda}{4\pi r_{p,n}}\gamma_{p}\mathrm{e}^{\mathrm{j}k_0 r_{p,n}},
	\label{hn}
\end{equation}
where $k_0 = 2\pi/\lambda$ denotes the wavenumber; {$r_{p,n} = \lVert\mathbf{s}_p - \mathbf{t}_n\rVert$ is the distance between user $p$ and antenna $n$ with $\mathbf{s}_p = [s_{p,x},s_{p,y},s_{p,z}]^{\mathrm{T}}$ and $\mathbf{t}_n = [t_{n,x}, t_{n,y}, t_{n,z}]^{\mathrm{T}}$ indicating their respective coordinates;} $\gamma_{p} \sim \mathcal{N}(\gamma_{p}; 0,1)$ denotes the complex path gain for user $p$\footnote{{In practical propagation environments, mmWave signals are influenced not only by free-space loss but also by the transmission medium. Consequently, the path loss may fluctuate around the free-space path loss. To provide flexible path loss modeling, we introduce the random variable $\gamma_p$ in (\ref{hn}). It is noteworthy that when setting $\gamma_p = 1$, (\ref{hn}) reduces to the result in \cite{Non_uniform}. Additionally, a similar modeling method is also presented in Eq. (6) of \cite{PolarCS}, where $l=1$ corresponds to the LoS path.}}. Furthermore, the channel vector $\mathbf{h}_p\in \mathbb{C}^{N\times 1}$ for user $p$ can be expressed as
\begin{equation}
		\mathbf{h}_p = \frac{\lambda}{4\pi} \gamma_{p}\mathbf{b}(\mathbf{r}_p),
		\label{hk}
\end{equation}
where $\mathbf{r}_p = [r_{p,1}, r_{p,2},\cdots, r_{p,N}]^{\mathrm{T}}$ denotes the distance vector between BS and user $p$; $\mathbf{b}(\mathbf{r}_p)$ is the array response vector with non-uniform spherical wavefront, denoted by
\begin{equation}
	\mathbf{b}(\mathbf{r}_p) = \left[\frac{\mathrm{e}^{\mathrm{j}k_0 r_{p,1}}}{r_{p,1}}, \frac{\mathrm{e}^{\mathrm{j}k_0 r_{p,2}}}{r_{p,2}}, \cdots,\frac{\mathrm{e}^{\mathrm{j}k_0 r_{p,N}}}{r_{p,N}}\right]^{\mathrm{T}}.
\end{equation} 
	
{Unfortunately, due to the nonlinear phase variations among antennas and the tight coupling between array response and distance, it is intractable to analyze the properties of the near-field channels directly from (\ref{hk}).} To address this, we employ the plane wave decomposition method presented in \cite{LoS_Angular, Weyl}. As a result, (\ref{hk}) can be expressed as a superposition of plane wave components with different spatial frequencies, i.e.,
\begin{equation}
	\mathbf{h}_p \approx \sqrt{N}\sum_{l =1}^{L}c_{p,l}\mathbf{a}(\xi_{l}),
	\label{hm2}
\end{equation}
where $c_{p,l}$ represents the complex gain of the $l$-th plane wave component for user $p$. The set encompassing all discrete spatial frequency components corresponding to different plane waves is given by
	\begin{equation}
		\begin{aligned}
			\mathcal{S} &= \left\{\xi \in \mathbb{Z}\Big| -k_0 \le \varDelta \xi < k_0 \right\},
		\end{aligned}
		\label{region}
	\end{equation}
where $\varDelta = {2\pi}/{(SD)}$ denotes the sampling spacing of spatial frequency; $S$ and $\xi_{l}$ indicate the oversampling factor and the $l$-th entry of $\mathcal{S}$, respectively; $L$ denotes the coordinate of $\mathcal{S}$. To illustrate the decomposition clearly, consider an example with the following parameters: $\lambda = 0.01$m, $N=256$, $D\approx128\lambda$, and $S=1$. According to (\ref{region}), we have $\mathcal{S} = \left\{-128,-127, \cdots, 126, 127\right\}$ and $L = 256$. The vector $\mathbf{a}(\xi_l) \in \mathbb{C}^{N\times 1}$ represents the array response vector corresponding to the $l$-th plane wave component, where its $n$-th entry is given by $[\mathbf{a}(\xi_{l})]_n = \exp\left(\mathrm{j}\varDelta\xi_{l}t_{n,x}\right)/{\sqrt{N}}$.
Moreover, (\ref{hm2}) can be expressed more succinctly in matrix form as
{\begin{equation}
	\mathbf{h}_p = {\mathbf{F}}\mathbf{h}_{a,p},
	\label{hm_matrix}
\end{equation}
where $\mathbf{h}_{a,p} = [c_{p,1}, c_{p,2}, \cdots, c_{p,L}]^{\mathrm{T}} \in \mathbb{C}^{L \times 1}$ denotes the wavenumber-domain channel vector for user $p$;} ${\mathbf{F}} \in \mathbb{C}^{N \times L} = [\mathbf{a}(\xi_1), \mathbf{a}(\xi_2), \cdots, \mathbf{a}(\xi_L)]$ is composed of stacked array response vectors and is semi-unitary. Consequently, $\mathbf{F}$ can be regarded as the wavenumber-domain codebook for the near-field channels.
	
According to (\ref{hm2}), the near-field channel with a spherical wavefront can be accurately characterized by $L$ plane wave components. It is crucial to note that, owing to the limited array size, only a few spatial frequency components in $\mathcal{S}$ significantly contribute to the near-field channels. These significant components typically concentrate within a specific range of spatial frequencies. To determine the range of significant spatial frequency components, the effective spatial bandwidth for user $p$ is defined as \cite{LoS_Angular}
\begin{equation}
	B_{p,e} = k_0\left(\max_{\mathbf{t}_n\in \mathcal{L}}\hat{\mathbf{d}}^{\mathrm{T}}(\mathbf{t}_n,\mathbf{s}_p)\hat{\mathbf{x}} - \min_{\mathbf{t}_n \in \mathcal{L}}\hat{\mathbf{d}}^{\mathrm{T}}(\mathbf{t}_n, \mathbf{s}_p)\hat{\mathbf{x}} \right), 
	\label{saptialBand}
\end{equation}
where $\hat{\mathbf{d}}(\mathbf{t}_n,\mathbf{s}_p) = (\mathbf{t}_n-\mathbf{s}_p)/\lVert \mathbf{t}_n-\mathbf{s}_p \rVert$ denotes the unit vector from $\mathbf{t}_n$ to $\mathbf{s}_p$; $\hat{\mathbf{x}}$ is the unit vector parallel to the ULA; $\mathcal{L}$ is the linear region spanned by the ULA. According to (\ref{saptialBand}), we can obtain the number of effective spatial frequency components for the $p$-th user as
\begin{equation}
	L_{p,e} \approx \lceil {B_{p,e}}/{\varDelta} \rceil .
	\label{S_e}
\end{equation}
where $ \lceil \cdot \rceil$ is to take the ceiling.

\begin{remark}
	{In this paper, various domains concerning near-field channels are encompassed, namely the antenna domain, wavenumber domain, and polar domain. To provide a clear understanding of these domains, we elaborate on their differences and connections as follows. The antenna-domain channel incorporates the channel responses of all antenna elements, as illustrated in (\ref{hk}).
	The wavenumber channel is derived through a linear transformation of the antenna channel, as depicted in (\ref{hm_matrix}). It reflects the significance of different spatial frequency components in accurately representing the antenna-domain channels. Due to the limited aperture size, the wavenumber-domain channel is typically sparse \cite{LoS_Angular}.
	Similarly, the polar-domain channel is also a linear transformation of the antenna-domain channel. Unlike the wavenumber-domain representation, which solely considers the sampling in the angular domain, the polar-domain representation simultaneously accounts for the sampling of angular and distance information. Consequently, the dimension of the polar domain transformation matrix is much larger than that of the wavenumber domain.}
\end{remark}
\vspace{-1em}
\subsection{Effects of SnS Property} 
\begin{figure}
	\centering
	\subfigure[Simulation Scenarios]{
		\includegraphics[width=0.16\textwidth]{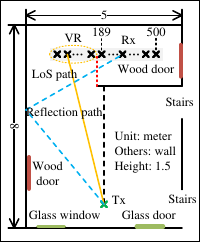}
		\label{SnSFig1}
	}
	\subfigure[Received power across elements]{
		\includegraphics[width=0.25\textwidth]{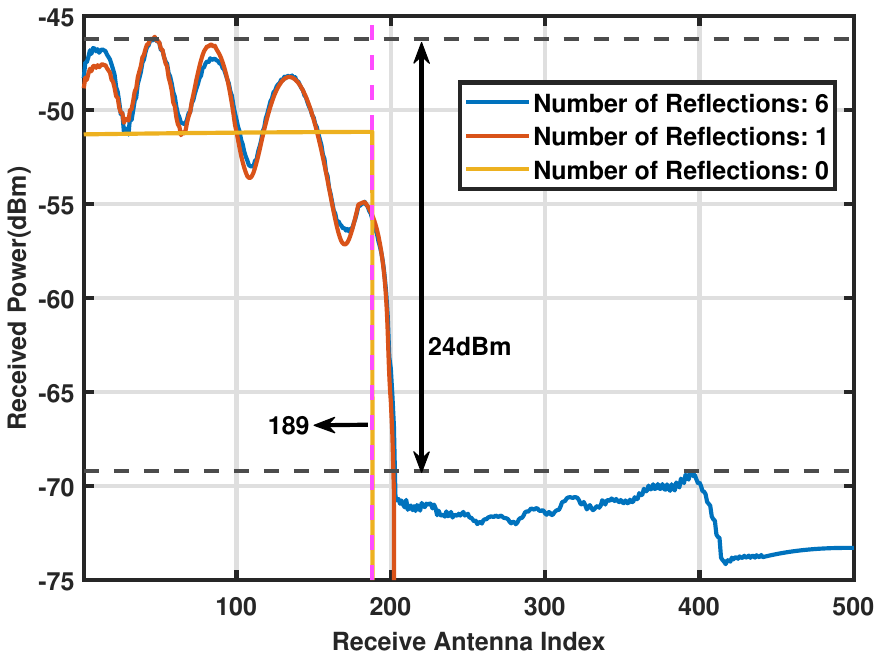}
		\label{SnSFig2}
	}
	\caption{SnS property simulations based on Rencom Wireless Insite.}
\end{figure}
	
{Another significant characteristic of XL-MIMO channels is the SnS property. In the LoS scenario, the SnS property primarily arises from obstacles that partially block the LoS propagation paths between users and antenna elements. To quantitatively characterize spatial non-stationarity, we employ VR-based channel modeling \cite{SnS1,SnS3}. Since obstacles may vary in size and appear at unpredictable locations, we do not explicitly model them in the paper. For simplification, similar to \cite{HanYu}, the VR is directly modeled as a random subarray of the array.}
Assume $\boldsymbol{\phi}_p \in \left\{1, 2, \cdots, N\right\}$ represent the set of visible antenna elements for user $p$. Thus, the VR indicator vector is denoted as $\boldsymbol{\alpha}_p \in \mathbb{Z}^{N\times 1}$, where its $n$-th element is given by
\begin{equation}
	{\alpha}_{p,n} = \left\{
	\begin{aligned}
		1&,\quad \text{if}\ n \in \boldsymbol {\phi}_p, \\
		0&,\quad \text{otherwise},\\
	\end{aligned}
	\right.
\end{equation}
with $\psi_p = {\left|\boldsymbol {\phi}_p \right|}/{N}$ indicating the proportion of array elements visible to the $p$-th user. 
Then, the SnS channel $\mathbf{x}_p$ between the BS and the $p$-th user can be written as
	\begin{equation}
		\begin{aligned}
			\mathbf{x}_p =\boldsymbol{\alpha}_p \odot \mathbf{h}_p=  \mathrm{diag}(\boldsymbol{\alpha}_p)\mathbf{h}_p,
		\end{aligned}
		\label{non_h}
	\end{equation}
where $\odot$ denotes the Hadamard product.

With the introduction of the SnS property, the spatially stationary codebook $\mathbf{F}$ becomes mismatched. Instead, the SnS codebook $\mathbf{F}^{\mathrm{SnS}}_p = \mathrm{diag}(\boldsymbol{\alpha}_p)\mathbf{F}$ should be utilized.
Consequently, the SnS channel in (\ref{non_h}) can be expressed as
\begin{equation}
	\begin{aligned}
		\mathbf{x}_p =\mathbf{F}^{\mathrm{SnS}}_p\mathbf{h}_{a,p}^{\mathrm{SnS}},
	\end{aligned}
	\label{non_h2}
\end{equation}
where $\mathbf{h}_{a,p}^{\mathrm{SnS}}$ denotes the wavenumber-domain channel vector for user $p$.

To empirically validate the model in (\ref{non_h}), akin to the ray-tracing simulations in \cite{SnS1} based on uniform circular arrays, we conduct simulations of the SnS property using a ULA.
The simulations are carried out in a 5m$\times$8m indoor room, as depicted in Fig. \ref{SnSFig1}. The transmitter and receiver consist of omnidirectional antennas positioned at a height of 1.5m above the ground. A narrow-band sinusoidal waveform at a frequency of 28 GHz is employed.
To simulate the SnS property, some of the receive array elements are obstructed by a wall, resulting in a subset of visible antenna elements denoted as $\boldsymbol{\phi} = \left\{1, 2, \cdots, 189\right\}$. Fig. \ref{SnSFig2} illustrates the received power across the elements for varying numbers of reflections.
When there is no reflection path, the received power of the array elements outside the VR is nearly zero. 
Even with reflection paths, the received power of the array elements outside the VR remains substantially lower than that of the array elements within the VR. This observation establishes that the VR-based channel model can effectively capture the SnS property with negligible error. This effectiveness is attributed to the fact that the received power of the array elements outside the VR is negligible.
	
\begin{figure}
		\centering
		\subfigure[$\psi_q = 0.7$]{
			\includegraphics[width=0.22\textwidth]{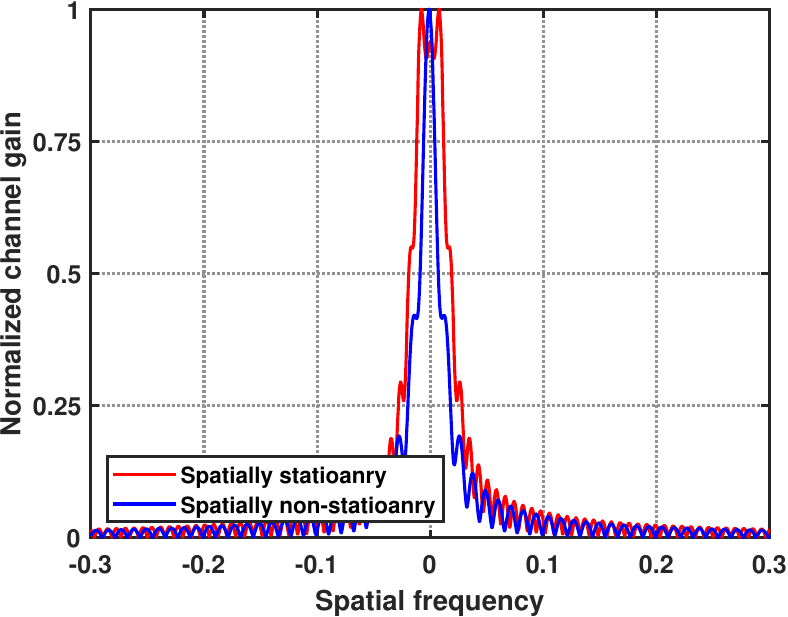}
			\label{case1}
		}
		\subfigure[$\psi_q =0.2$]{
			\includegraphics[width=0.22\textwidth]{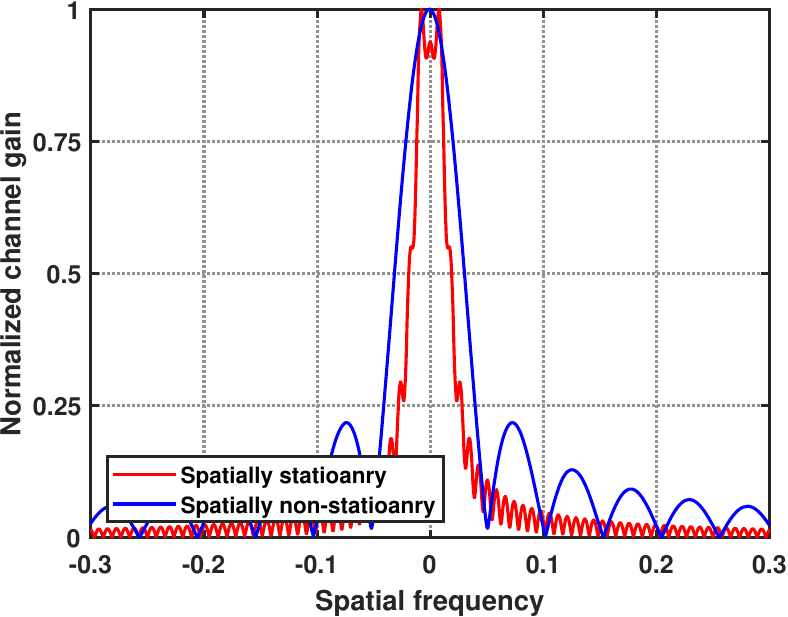}
			\label{case2}
		}
		\caption{The wavenumber-domain channel of spatial stationary and non-stationary scenarios.}
		\label{wavenumber_domain}
	\end{figure}

To assess the influence of the SnS property on wavenumber-domain sparsity, we examine the wavenumber-domain characteristics in both spatially stationary and SnS scenarios. The parameters are set as $N=256$, $\lambda=3\times10^{-3}$m, $\psi= 0.7$ and $ 0.2$. The distance between the center of the ULA and the user is fixed at $10$m. From Fig. \ref{wavenumber_domain}, it is evident that the spatially stationary and SnS channels exhibit distinct sparsity characteristics. The distinctions in the wavenumber-domain channels can be elaborated upon from the following two aspects:
\begin{enumerate}[1)]
	\item The existence of the SnS property implies that only a portion of the array can capture the electromagnetic wave radiated by the user. This is equivalent to reducing the effective receiving size of the array between the BS and the user. As described in (\ref{saptialBand}), when the effective size of the array is reduced, the effective spatial bandwidth between the user and the array will also correspondingly diminish.
	\item The decrease in the effective array size also results in a more significant spectrum (spatial frequency) leakage phenomenon, where the beamwidth corresponding to each spatial frequency component will be extended\cite{Energy_Spread}. 
\end{enumerate} 
	Therefore, the block sparsity of SnS XL-MIMO channels in the wavenumber domain results from the synthesis of two factors. Specifically, when the VR is relatively large, the predominant influence stems from the reduction in effective spatial frequency components. Consequently, the wavenumber-domain channel exhibits more pronounced sparsity, as illustrated in Fig. \ref{case1}. Conversely, when the VR is relatively small, the spectrum leakage phenomenon takes precedence, leading to an increase in significant spatial frequency components, as depicted in Fig. \ref{case2}.

\section{Problem Reformulation and Solutions}
\label{section3}
In this section, we first introduce the joint VR detection and CE protocol. We then formulate VR detection as a Bayesian inference problem, efficiently solved using a structured MP scheme in the antenna domain. Subsequently, leveraging the obtained VR information and the wavenumber-domain sparsity, we derive the channel estimation as a sparse signal recovery problem, solved through an OMP-based method.
\vspace{-1em}
\subsection{Joint VR Detection and Channel Estimation Protocol}
\begin{figure}
	\centering
	\includegraphics[width=0.45\textwidth]{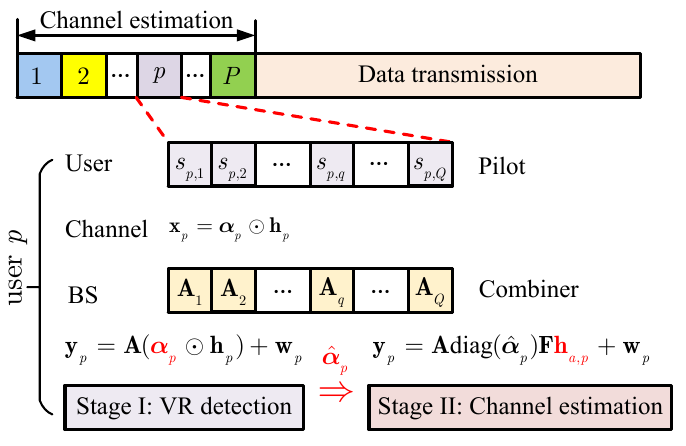}
	\caption{The proposed two-stage CE protocol}
	\label{Protocol}
\end{figure}

Due to the channel reciprocity of the TDD mode, we focus on formulating the VR detection and CE in the uplink. The downlink channels can be straightforwardly obtained based on the results of the uplink. We adopt the commonly employed orthogonal pilot transmission strategy \cite{orthogonal_1, orthogonal_2}, where different users transmit mutually orthogonal pilot sequences (e.g., orthogonal time or frequency resources) to the BS. Consequently, the channel estimation for each user becomes an independent process, allowing for sequential estimation.  
Let $s_{p,q}$ denote the transmit pilot of the $p$-th user in time slot $q$, as illustrated in Fig.~\ref{Protocol}. At this point, the received signal $\mathbf{y}_{p,q} \in \mathbb{C}^{N{\mathrm{rf}} \times 1}$ in RF chains can be expressed as
	\begin{equation}
		\mathbf{y}_{p,q} = \mathbf{A}_q\mathbf{x}_p s_{p,q} + \mathbf{A}_q\mathbf{n}_{p,q},
	\end{equation}
where $\mathbf{x}_p$ is the antenna-domain channel vector given in (\ref{non_h});  $\mathbf{A}_q \in \mathbb{C}^{N_{\mathrm{rf}} \times N}$ and $\mathbf{n}_{p,q}\sim \mathcal{CN}(\mathbf{n}_{p,q}; 0, \sigma_{\mathrm{N}}^2\mathbf{I}_N) \in \mathbb{C}^{N\times1}$ denote the analog combiner matrix and the complex additional white Gaussian noise (AWGN) in the $q$-th time slot for user $p$, respectively. Without loss of generality, for any $p$ and $q$, we assume $s_{p,q} = 1$. Collecting all $Q$ pilot symbols, the overall received pilot sequence $\mathbf{y}_p = [\mathbf{y}_{p,1}^{\mathrm{T}},\mathbf{y}_{p,2}^{\mathrm{T}}, \cdots,\mathbf{y}_{p,Q}^\mathrm{T}]^\mathrm{T} \in \mathbb{C}^{QN_{\mathrm{rf}}\times1}$ for user $p$ can be denoted as 
	\begin{equation}
		\mathbf{y}_p = \mathbf{A}\mathbf{x}_p + \mathbf{w}_p,
		\label{pilot}
	\end{equation}
{where $\mathbf{A} = [\mathbf{A}_1^{\mathrm{T}}, \mathbf{A}_2^{\mathrm{T}}, \cdots, \mathbf{A}_Q^{\mathrm{T}}]^{\mathrm{T}}$ $\in \mathbb{C}^{M\times N}$ denotes the overall combiner matrix, which is composed of $M = QN_{\mathrm{rf}}$ randomly selected rows of the normalized DFT matrix, whose $(m,n)$-th entry is given by $\frac{1}{\sqrt{N}}\exp\left({\mathrm{j}\frac{2\pi(m-1)(n-1)}{N}} \right)$;} 
$\mathbf{w}_p = [\mathbf{n}^{\mathrm{T}}_{p,1}\mathbf{A}_1^{\mathrm{T}}, \mathbf{n}^{\mathrm{T}}_{p,2}\mathbf{A}_2^{\mathrm{T}},\cdots,\mathbf{n}^{\mathrm{T}}_{p,Q}\mathbf{A}_Q^{\mathrm{T}}]^{\mathrm{T}} \in \mathbb{C}^{M\times 1}$ denotes the effective noise. Since $\mathbf{A}$ is a semi-unitary matrix, i.e., $\mathbf{A}\mathbf{A}^{\mathrm{H}} = \mathbf{I}_M$, $\mathbf{w}_p$ is still Gaussian noise satisfying $\mathcal{CN}(\mathbf{w}_{p}; 0, \sigma^2_{\mathrm{N}}\mathbf{I}_M)$. {Since the channel estimation for each user is independent, we consider user $p$ as an example and other users can follow a similar procedure. For brevity, we omit the subscript $p$ in the following part of this subsection.}

Recently, various schemes have recently been explored for VR detection and CE \cite{HanYu, Hanyu2, Bayesian3, Bayesian1}. However, these existing methods face two limitations. 1) The spatial correlation among adjacent antennas do not be exploited, where the non-zero elements of $\boldsymbol{\alpha}$ tend to concentrate within a specific portion of the array. 
2) The correlation between antennas in the wavenumber domain cannot be fully exploited.
These limitations could lead to degradation in estimation performance. Consequently, to enhance the VR detection and CE performance, we propose to simultaneously exploit spatial correlation of the VR in the antenna domain and the wavenumber-domain sparsity in this work. By utilizing the representation in (\ref{non_h2}), the overall received signal can be rewritten as
	\begin{equation}
		\mathbf{y} = \mathbf{A}\mathbf{x}+\mathbf{w} =\mathbf{A}\mathbf{F}^{\mathrm{SnS}}\mathbf{h}_{a}^{\mathrm{SnS}} + \mathbf{w},
		\label{pilot3}
	\end{equation} 
where $\mathbf{F}^{\mathrm{SnS}}$ represents the SnS wavenumber-domain codebook; $\mathbf{A}$ denotes the overall analog combiner matrix.

	
Our primary objective is to accurately estimate $\boldsymbol{\alpha}$ and $\mathbf{h}_a^{\mathrm{SnS}}$ based on the observation vector $\mathbf{y}$. However, this estimation task encounters two challenges. Firstly, the number of RF chains is considerably smaller than the number of antennas, preventing the BS from simultaneously capturing signals from all antennas. Secondly, the estimation of $\boldsymbol{\alpha}$ and $\mathbf{h}_a$ takes place in different domains, posing difficulties for concurrent estimation. To solve these problems, we propose to transform the concurrent estimation into sequential estimation, and 
a two-stage estimation scheme is developed. As shown in Fig.~\ref{Protocol}, in the first stage, our focus is on detecting the user's VR  based on the received signal model in (\ref{pilot}). In the second stage, based on the received signal model in (\ref{pilot3}), the channel estimation is derived as a sparse signal recovery problem.
\vspace{-1em}
\subsection{Stage \uppercase\expandafter{\romannumeral1}: VR Detection in Antenna Domain}
	Due to the SnS property of XL-MIMO channels, we assume that the channel coefficients conditioned on the VR are independent and identically distributed as Bernoulli Gaussian \cite{Bayesian1, Bayesian3}, which is given by
	\begin{equation}
		p(\mathbf{x}|\boldsymbol{\alpha}) = \prod_{n=1}^{N}p(x_n|\alpha_n),
		\label{pxa}
	\end{equation} 
	with $p(x_n|\alpha_n) = (1-\alpha_n)\delta(x_n) + \alpha_n\mathcal{CN}(x_n;0,\sigma^2)$, where $\delta(\cdot)$ denotes the Dirac delta function, which evaluates to 1 when $x=0$ and 0 otherwise; $x_n$ is the channel coefficient between the $n$-th antenna and the user; $\alpha_n$ denotes the corresponding visibility indicator. It is evident that when the antenna element $n$ is invisible to the user, indicated by $\alpha_n=0$, the channel coefficient $x_n$ is zero. Otherwise, when $\alpha_n=1$, the channel coefficient $x_n$ follows a Gaussian distribution with mean zero and variance $\sigma^2$.
	
{To capture this spatial correlation of the VR, $\boldsymbol{\alpha}$ can be modeled as one-order Markov chain}\footnote{{In practical XL-MIMO systems, each antenna element may be related to multiple nearby antennas.  However, for the sake of simplicity, we limit our analysis to a one-order Markov chain to capture the spatial correlations of adjacent antennas in this paper. In our subsequent work, we are considering the adoption of a Markov random field for more accurate modeling.}}, which is given by 
	\begin{equation}
		p(\boldsymbol{\alpha}) = p(\alpha_1) \prod_{n=2}^N p(\alpha_n|\alpha_{n-1}),
		\label{pa}
	\end{equation}
where the steady state probability of the Markov chain is given by $p(\alpha_n = 1) = \psi$. The transition probability of the Markov chain $p(\alpha_n|\alpha_{n-1})$, $n = 2,3,\cdots,N$, is given by
	\begin{equation}
		p(\alpha_n|\alpha_{n-1}) = \left\{
		\begin{aligned}
			(1-p_{01})^{1-\alpha_n}p_{01}^{\alpha_n}&,\quad \alpha_{n-1} = 0, \\
			p_{10}^{1-\alpha_n} (1-p_{10})^{\alpha_n}&,\quad \alpha_{n-1} = 1,\\
		\end{aligned}
		\right.
		\label{trans}
	\end{equation}
where $p_{01} = p(\alpha_n = 0|\alpha_{n-1} = 1)$ and the other three transition probabilities are defined similarly. With the steady-state assumption, the Markov chain can be completely characterized by two parameters $\psi$ and $p_{10}$. The other three transition probabilities can be easily obtained as $p_{01} = \psi p_{10}/(1-\psi)$, $p_{00} = 1-p_{01}$, and $p_{11} = 1-p_{10}$.
	
Combining the prior distribution provided in (\ref{pxa}) and (\ref{pa}), the probabilistic signal model describing the problem in (\ref{pilot}) can be derived as
\begin{equation}
	p(\mathbf{y}, \mathbf{x},\boldsymbol{\alpha}) =p(\mathbf{y}|\mathbf{x},\boldsymbol{\alpha})p(\mathbf{x}|\boldsymbol{\alpha})p(\boldsymbol{\alpha}),
	\label{prob_model}
\end{equation}
where $p(\mathbf{y}|\mathbf{x},\boldsymbol{\alpha}) \propto \mathcal{CN}(\mathbf{y}; \mathbf{Ax}, \sigma^2_{\mathrm{N}})$ due to the AWGN observation model.

Based on the probabilistic model provided in (\ref{prob_model}), the optimal detector of VR in terms of the minimum mean square error (MMSE) principle can be formulated as maximum a posterior (MAP) estimation problem, i.e., 
\begin{equation}
	\hat{\boldsymbol{\alpha}} \propto \mathop{\arg\max}\limits_{\boldsymbol{\alpha}} p(\mathbf{y}|\mathbf{x},\boldsymbol{\alpha})p(\mathbf{x}|\boldsymbol{\alpha})p(\boldsymbol{\alpha}).
	\label{posterior}
\end{equation}

Since the number of antenna elements in XL-MIMO systems is extremely large, it is computationally intractable to obtain the exact solution of (\ref{posterior}). {Recently, approximate MP techniques \cite{gamp_rangan, gamp, MP_Contributions} with low complexity have been extensively utilized for the MAP estimation problem.
Specifically, a Turbo based MP scheme has been widely adopted for sparse signal recovery\cite{Turbo_Ma1,Turbo_Ma2}, channel estimation\cite{HMMLiuAn, HMMLiuAn2}, and MIMO detection\cite{Turbo_Jin1,Turbo_Jin2}. 
Motivated by the superiority of Turbo based MP in MAP estimation, we utilize the structured MP framework to obtain the VR. 
Specifically, the VRDO-MP scheme decouples the problem (\ref{posterior}) into two MMSE estimators, as shown in Fig.~\ref{Factor},  where the likelihood and prior are represented as factor nodes (gray rectangles), while the random variables are represented as variable nodes (blank circles), and the gray circles indicate the observed values.
\begin{itemize}
 	\item \textbf{Module A: linear MMSE estimator}, where $\mathbf{x}$ is estimated based on the observation $\mathbf{y}$ and  and the extrinsic message from Module B;
 	\item \textbf{Module B: VR detector}, where $\boldsymbol{\mathbf{\alpha}}$ is estimated by combining the structured priors $p(\mathbf{x}|\boldsymbol{\alpha})$, $p(\boldsymbol{\alpha})$ and the extrinsic message from Module A. 
\end{itemize}
\begin{figure*}
	\centering
	\includegraphics[width=0.9\textwidth]{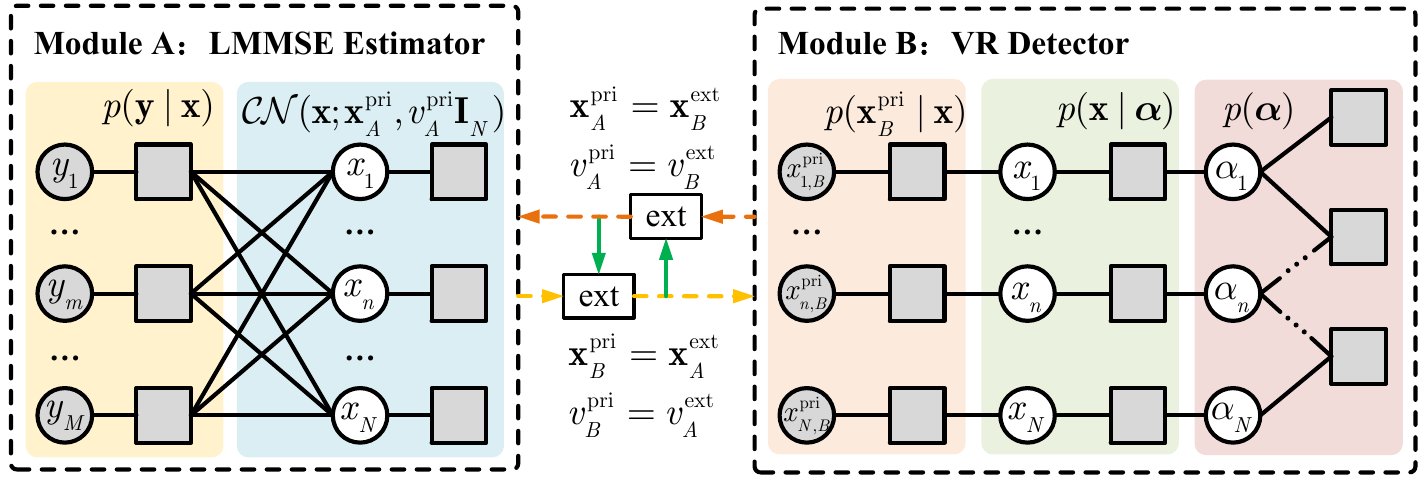}
	\caption{The block diagram of the VRDO-MP scheme.}
	\label{Factor}
\end{figure*}
	Since the information of the VR is implicit in $\mathbf{x}$, we utilize $\mathbf{x}$ as a bridge for MP between Module A and Module B.
\subsubsection{LMMSE Estimator} 
In Module A, the antenna-domain channel $\mathbf{x}$ is estimated based on the received pilot vector $\mathbf{y}$ and the prior distribution $\mathcal{CN}(\mathbf{x}; \mathbf{x}_{A}^{\mathrm{pri}}, v_{A}^{\mathrm{pri}}\mathbf{I}_{N})$, where $\mathbf{x}_{A}^{\mathrm{pri}}$ and $v_{A}^{\mathrm{pri}}$ are extrinsic inputs from Module B, which is elaborated later. Using the LMMSE rule \cite{LMMSE}, the posterior mean and the covariance of $\mathbf{x}$ are respectively derived as
\begin{align}
	\notag \mathbf{x}_{A}^{\mathrm{post}} 
	&= \mathbf{x}_{A}^{\mathrm{pri}} + v_{A}^{\mathrm{pri}}\mathbf{A}^{\mathrm{H}}\left(v_{A}^{\mathrm{pri}}\mathbf{A}\mathbf{A}^{\mathrm{H}} +  \sigma_N^2\mathbf{I}_M \right)^{-1}(\mathbf{y}-\mathbf{A}\mathbf{x}_{A}^{\mathrm{pri}}) \\
	\label{xapost} &\overset{(a)}{=} \mathbf{x}_{A}^{\mathrm{pri}} + \frac{v_{A}^{\mathrm{pri}}}{v_{A}^{\mathrm{pri}} + \sigma_N^2}\mathbf{A}^{\mathrm{H}}(\mathbf{y}-\mathbf{A}\mathbf{x}_{A}^{\mathrm{pri}}),\\
	\notag \mathbf{V}_{A}^{\mathrm{post}} &= v_{A}^{\mathrm{pri}}{\mathbf{I}_N}-(v_{A}^{\mathrm{pri}})^2\mathbf{A}^{\mathrm{H}}\left(v_{A}^{\mathrm{pri}} \mathbf{A}\mathbf{A}^{\mathrm{H}} +  \sigma_N^2\mathbf{I}_M \right)^{-1}\mathbf{A}\\  
	\label{vapost} &\overset{(b)}{=} \left(v_{A}^{\mathrm{pri}} -\frac{M}{N} \frac{(v_{n,A}^{\mathrm{pri}})^2}{v_{n,A}^{\mathrm{pri}} + \sigma_N^2}\right) \mathbf{I}_N.
\end{align}
{where $(a)$ and $(b)$ are obtained due to the unitary property of $\mathbf{A}$ that yields $\mathbf{A}^{\mathrm{H}}\mathbf{A} = (M/N)\mathbf{I}_N$ and $\mathbf{A}{\mathbf{A}^\mathrm{H}} =  \mathbf{I}_M$. The property effectively avoids the matrix inverse operation in the LMMSE estimator.}

Then, the extrinsic outputs of Module A are respectively given by \cite{Bayesian1,HMMLiuAn2}
\begin{align}
	v_{A}^{\mathrm{ext}} = \frac{v_{A}^{\mathrm{pri}}v_{A}^{\mathrm{post}}}{v_{A}^{\mathrm{pri}}-v_{A}^{\mathrm{post}}}, \,
	\label{vaext} \mathbf{x}_{A}^{\mathrm{ext}} = v_{A}^{\mathrm{ext}}\left(\frac{\mathbf{x}_{A}^{\mathrm{post}}}{v_{A}^{\mathrm{post}}} - \frac{\mathbf{x}_{A}^{\mathrm{pri}}}{v_{A}^{\mathrm{pri}}}\right),
\end{align}
which serve as the input information of Module B, i.e., $\mathbf{x}_{B}^{\mathrm{pri}} = \mathbf{x}_{A}^{\mathrm{ext}}$ and $
{v}_{B}^{\mathrm{pri}} = v_{A}^{\mathrm{ext}}$.
\subsubsection{VR Detector} {Assume the input messages of Module B, $\mathbf{x}_{B}^{\mathrm{pri}}$, can be approximated as an additive white Gaussian noise observation\footnote{{In practical XL-MIMO systems, the approximation might be far from the Gaussian probability density function. In the future, we will further consider utilize both the model and data driven method to obtain more accurate message for the VR detector.}} \cite{Turbo_Ma1},} i.e.,
\begin{equation}
	\mathbf{x}_{B}^{\mathrm{pri}} = \mathbf{x} + \mathbf{z},
\end{equation}
where $\mathbf{z} \sim \mathcal{CN}(\mathbf{z}; 0, {v}_{B}^{\mathrm{pri}}\mathbf{I}_N)$ is independent of $\mathbf{x}$, and ${v}_{B}^{\mathrm{pri}}$ is the extrinsic variance from Module A. Therefore, the observation variable nodes $\{y_m \}$ of the factor graph in Fig. \ref{Factor} are replaced by $\{ {x}_{n, B}^{\mathrm{pri}} \}$, and the factor nodes $\{ p(y_m|\mathbf{x})\}$ are replaced by $\{p(x_{n,B}^{\mathrm{pri}}|x_n) \}$. Therefore, we have
\begin{equation}
	p(x_{n,B}^{\mathrm{pri}}|x_n) = \mathcal{CN}(x_{n,B}^{\mathrm{pri}}; x_n, v_B^{\mathrm{pri}}).
	\label{prior1}
\end{equation}

In the following, we will derive the MP procedure step by step in Module B according to the sum-product rules. The definitions of the involved messages for Module B are summarized in Table \ref{Tab1}.

\begin{table*}
\renewcommand\arraystretch{1.5}
\begin{center}
	\caption{Message Definition for Module B}
	\begin{tabular}{|l|l|l|}
		\hline
		\multicolumn{1}{|c|}{Notation} & \multicolumn{1}{c|}{Definition} & \multicolumn{1}{c|}{Distribution} \\
		\hline
		$\nu_{p(x_{n,B}^{\mathrm{pri}}|x_n) \rightarrow x_n} (x_n)$ & message from $p(x_{n,B}^{\mathrm{pri}}|x_n)$ to $x_n$ & $\mathcal{CN}(x_n; x_{n,B}^{\mathrm{pri}}, v_{n,B}^{\mathrm{pri}})$\\
		\hline
		$\nu_{x_n \rightarrow p(x_n|\alpha_n)}(x_n)$ & message from $x_n$ to $p(x_n|\alpha_n)$ & $\mathcal{CN}(x_n; x_{n,B}^{\mathrm{pri}}, v_{n,B}^{\mathrm{pri}})$\\
		\hline
		$\nu_{p(x_n|\alpha_n) \rightarrow \alpha_n}(\alpha_n)$ & message from $p(x_n|\alpha_n)$ to $\alpha_n$ & $\pi_n^{\mathrm{out}} \delta(1-\alpha_n) + (1-\pi_n^{\mathrm{out}})\delta(\alpha_n)$\\
		\hline
		$\nu_{\alpha_n \rightarrow p(\alpha_{n+1}|\alpha_{n})}$ & message from $\alpha_n$ to $p(\alpha_{n+1}|\alpha_{n})$ & $\lambda^f_n\pi_n^{\mathrm{out}} \delta(1-\alpha_{n})+(1-\lambda^f_n)(1-\pi_n^{\mathrm{out}})\delta(\alpha_{n})$\\
		\hline
		$\nu_{p(\alpha_{n}|\alpha_{n-1}) \rightarrow \alpha_n}(\alpha_{n})$&message from $p(\alpha_{n}|\alpha_{n-1})$ to $\alpha_n$ & \makecell[l]{$(1-\lambda^f_{n-1})(1-\pi_{n-1}^{\mathrm{out}}) p(\alpha_{n}|\alpha_{n-1} = 0)$ \\$+\lambda^f_{n-1} \pi_{n-1}^{\mathrm{out}} p(\alpha_{n}|\alpha_{n-1} = 1)$}\\
		\hline
		$\nu_{\alpha_n \rightarrow p(\alpha_n|\alpha_{n-1})}(\alpha_n)$& message from $\alpha_n$ to $p(\alpha_{n}|\alpha_{n-1})$ & $\lambda^b_{n}\pi_n^{\mathrm{out}} \delta(1-\alpha_{n})+(1-\lambda^b_{n})(1-\pi_{n}^{\mathrm{out}})\delta(\alpha_{n})$\\
		\hline
		$\nu_{p(\alpha_{n+1}|\alpha_{n}) \rightarrow \alpha_{n}}(\alpha_{n})$& message from $p(\alpha_{n+1}|\alpha_{n})$ to $\alpha_{n}$ & \makecell[l]{$(1-\lambda^b_{n+1})(1-\pi_{n+1}^{\mathrm{out}}) p(\alpha_{n+1}=0|\alpha_{n})$\\ $+ \lambda^b_{n+1}\pi_{n+1}^{\mathrm{out}} p(\alpha_{n+1}=1|\alpha_{n})$} \\
		\hline
		$\nu_{\alpha_n \rightarrow p(x_n|\alpha_n)}(\alpha_n)$ &  message from $\alpha_n$ to $p(x_n|\alpha_n)$ & $\pi^{\mathrm{in}}_n \delta(1-\alpha_{n}) + (1-\pi^{\mathrm{in}}_n) \delta(\alpha_{n})$\\
		\hline
		$\nu_{p(x_n|\alpha_n) \rightarrow x_n}(x_n)$&  message from $p(x_n|\alpha_n)$ to $x_n$ & $\pi^{\mathrm{in}}_n\mathcal{CN}(x_n; 0, \sigma^2) + (1-\pi^{\mathrm{in}}_n)\delta(x_n)$\\
		\hline
		$b(\alpha_{n})$& approximate posterior distribution of $\alpha_n$ & $\lambda_n^{\mathrm{post}}\delta(1-\alpha_n) + (1-\lambda_n^{\mathrm{post}})\delta(\alpha_n)$\\
		\hline
	\end{tabular}
\label{Tab1}  
\end{center}
\end{table*}

We first consider the forward passing, i.e., message passing from left to right. Since the variable node $x_n$ is only connected to the factor nodes $p(x_{n,B}^{\mathrm{pri}}|x_n)$ and $p(x_n|\alpha_n)$, according to the previous assumption in (\ref{prior1}), the message from  $p(x_{n,B}^{\mathrm{pri}}|x_n)$ to  $x_n$ and from  $x_n$ to  $p(x_n|\alpha_n)$ can be respectively given by
\begin{align}
\label{vpx1} \nu_{p(x_{n,B}^{\mathrm{pri}}|x_n) \rightarrow x_n} (x_n) = \mathcal{CN}(x_n; x_{n,B}^{\mathrm{pri}}, v_{n,B}^{\mathrm{pri}}),\\
\label{vxp} \nu_{x_n \rightarrow p(x_n|\alpha_n)} (x_n) = \mathcal{CN}(x_n; x_{n,B}^{\mathrm{pri}}, v_{n,B}^{\mathrm{pri}}),
\end{align}

Utilizing (\ref{vxp}) and the sum-product rule, the message from $p(x_n|\alpha_n)$ to $\alpha_n$ can be derived as
\begin{equation}
\begin{aligned}
	&\nu_{p(x_n|\alpha_n) \rightarrow \alpha_n}(\alpha_n)\\
	=& \int \nu_{x_n \rightarrow p(x_n|\alpha_n)}(x_n) p(x_n|\alpha_n) \mathrm{d}x_n\\
	\propto& \pi_n^{\mathrm{out}} \delta(1-\alpha_n) + (1-\pi_n^{\mathrm{out}})\delta(\alpha_n),
\end{aligned}
\label{vpa1}
\end{equation}
where $\pi_n^{\mathrm{out}} ={d_n}/({\mathcal{CN}(0; x_{n,B}^{\mathrm{pri}}, v_{n,B}^{\mathrm{pri}})+d_n})$
with $d_n \triangleq \mathcal{CN}(0;x_{n,B}^{\mathrm{pri}}, \sigma^2+v_{n,B}^{\mathrm{pri}})$.

Then, we derive the forward MP across the Markov chain. Denote $\psi^f_n$ as the message $p(\alpha_{n}|\alpha_{n-1})$ to $\alpha_n$. Thus,  for any $ n\ge2$, we have
\begin{equation}
	\nu_{p(\alpha_n|\alpha_{n-1}) \rightarrow \alpha_n}(\alpha_n) = \psi^f_n \delta(1-\alpha_{n})+(1-\psi^f_n)\delta(\alpha_{n}).
\end{equation} 
For $n=1$, we always have $\nu_{p(\alpha_n|\alpha_{n-1}) \rightarrow \alpha_n}(\alpha_n) = \psi\delta(1-\alpha_n)+(1-\psi)\delta(\alpha_n)$, since $p(\alpha_1)$ is only connected with $\alpha_1$ in the factor graph. Then, the message from variable node $\alpha_n$ to factor node $p(\alpha_{n+1}|\alpha_{n})$ can be formulated as
\begin{equation}
	\begin{aligned}
				&\nu_{\alpha_n \rightarrow p(\alpha_{n+1}|\alpha_{n})}(\alpha_n) \\= & \nu_{p(x_n|\alpha_n) \rightarrow \alpha_n}(\alpha_{n}) \nu_{p(\alpha_n|\alpha_{n-1}) \rightarrow \alpha_n}(\alpha_{n})\\
				= &\psi^f_n\pi_n^{\mathrm{out}} \delta(1-\alpha_{n})+(1-\psi^f_n)(1-\pi_n^{\mathrm{out}})\delta(\alpha_{n}).
			\end{aligned}
			\label{Mf1}
		\end{equation}
		
Similar to (\ref{Mf1}), we can obtain $\nu_{\alpha_{n-1} \rightarrow p(\alpha_{n}|\alpha_{n-1})}(\alpha_{n-1}) = \psi^f_{n-1}\pi_n^{\mathrm{out}} \delta(1-\alpha_{n-1})+(1-\psi^f_{n-1})(1-\pi_{n-1}^{\mathrm{out}})\delta(\alpha_{n-1})$. Thus, the message from factor node $p(\alpha_{n}|\alpha_{n-1})$ to variable node $\alpha_n$ is given by
		\begin{equation}
			\begin{aligned}
				&\nu_{p(\alpha_{n}|\alpha_{n-1}) \rightarrow \alpha_n}(\alpha_{n})  \\=& \int \nu_{\alpha_{n-1} \rightarrow p(\alpha_n|\alpha_{n-1})}(\alpha_{n-1}) p(\alpha_{n}|\alpha_{n-1}) \mathrm{d}\alpha_{n-1}\\
				=& (1-\psi^f_{n-1})(1-\pi_{n-1}^{\mathrm{out}}) p(\alpha_{n}|\alpha_{n-1} = 0)\\ +& \psi^f_{n-1} \pi_{n-1}^{\mathrm{out}} p(\alpha_{n}|\alpha_{n-1} = 1).
			\end{aligned}
			\label{a2pa}
		\end{equation}
		Combining (\ref{trans}) and (\ref{a2pa}), we can obtain
		\begin{equation}
			\begin{aligned}
				\psi^f_n = \frac{p_{01}(1-\psi^f_{n-1})(1-\pi_{n-1}^{\mathrm{out}})+ p_{11}\lambda^f_{n-1} \pi_{n-1}^{\mathrm{out}}}{(1-\psi^f_{n-1})(1-\pi_{n-1}^{\mathrm{out}})+ \psi^f_{n-1} \pi_{n-1}^{\mathrm{out}}}.
			\end{aligned} 
			\label{Mf2}
		\end{equation}
		
		Similar to the forward MP, for any $ n \le N-1$, denote $\psi^b_n$ as the backward MP of the Markov chain from $p(\alpha_{n+1}|\alpha_{n})$ to $\alpha_{n}$, i.e.,
		\begin{equation}
			\nu_{p(\alpha_{n+1}|\alpha_{n}) \rightarrow \alpha_{n}}(\alpha_{n}) = \psi^b_n \delta(1-\alpha_{n})+(1-\psi^f_n)\delta(\alpha_{n}),
		\end{equation}
		For $n=N$, we always set $\psi^b_N = 1/2$. Then, the message from $\alpha_n$ to $p(\alpha_{n}|\alpha_{n-1})$ is written as
		\begin{equation}
			\begin{aligned}
				&\nu_{\alpha_n \rightarrow p(\alpha_n|\alpha_{n-1})}(\alpha_n)\\
				=& \nu_{p(x_n|\alpha_n) \rightarrow \alpha_n}(\alpha_{n}) \nu_{p(\alpha_{n+1}|\alpha_{n}) \rightarrow \alpha_n}(\alpha_{n})\\
				=& \psi^b_{n}\pi_n^{\mathrm{out}} \delta(1-\alpha_{n})+(1-\psi^b_{n})(1-\pi_{n}^{\mathrm{out}})\delta(\alpha_{n}).
			\end{aligned}
			\label{Mb1}
		\end{equation}
		 Similar to (\ref{Mb1}), we can obtain $\nu_{\alpha_{n+1} \rightarrow p(\alpha_{n+1}|\alpha_{n})}(\alpha_{n+1}) = \psi^b_{n+1}\pi_{n+1}^{\mathrm{out}} \delta(1-\alpha_{n+1})+(1-\psi^b_{n+1})(1-\pi_{n+1}^{\mathrm{out}})\delta(\alpha_{n+1})$. Thus, the message from $p(\alpha_{n+1}|\alpha_{n})$ to $\alpha_{n}$ is given by
		\begin{equation}
			\begin{aligned}
				&\nu_{p(\alpha_{n+1}|\alpha_{n}) \rightarrow \alpha_{n}}(\alpha_{n})\\
				=& \int \nu_{\alpha_{n+1} \rightarrow p(\alpha_{n+1}|\alpha_{n})}(\alpha_{n+1}) p(\alpha_{n+1}|\alpha_{n}) \mathrm{d}\alpha_{n+1} \\
				=&(1-\psi^b_{n+1})(1-\pi_{n+1}^{\mathrm{out}}) p(\alpha_{n+1}=0|\alpha_{n})
				\\ +& \psi^b_{n+1}\pi_{n+1}^{\mathrm{out}} p(\alpha_{n+1}=1|\alpha_{n}). 
			\end{aligned}
			\label{pa2a}
		\end{equation}
		According to (\ref{trans}) and (\ref{pa2a}), the backward MP across the Markov chain can be obtained in (\ref{Mb2}) shown in the top of the next page.
		\begin{figure*}
			\normalsize
			\setcounter{MYtempeqncnt}{\value{equation}}
			\setcounter{equation}{34}
			\begin{align}
				\label{Mb2} \psi^b_n =& \frac{p_{10}(1-\psi^b_{n+1})(1-\pi_{n+1}^{\mathrm{out}})+p_{11}\psi^b_{n+1}\pi_{n+1}^{\mathrm{out}}}{(p_{10}+p_{00})(1-\psi^b_{n+1})(1-\pi_{n+1}^{\mathrm{out}})+(p_{11}+p_{01})\psi^b_{n+1}\pi_{n+1}^{\mathrm{out}}},\\
				\notag {b}(\alpha_{n}) \propto& \nu_{p(x_n|\alpha_n) \rightarrow \alpha_n}(\alpha_n) \nu_{p(\alpha_{n}|\alpha_{n-1}) \rightarrow \alpha_n}(\alpha_{n}) \nu_{p(\alpha_{n+1}|\alpha_{n}) \rightarrow \alpha_{n}}(\alpha_{n})\\
				\notag =&\pi_n^{\mathrm{out}}\psi^f_n \psi^b_n \delta(1-\alpha_n) + (1-\pi_n^{\mathrm{out}})(1-\psi^f_n)(1-\psi^b_n)\delta(\alpha_n)\\
				\label{belief} \propto& \psi_n^{\mathrm{post}}\delta(1-\alpha_n) + (1-\psi_n^{\mathrm{post}})\delta(\alpha_n),
			\end{align}
			\hrulefill
			\vspace*{4pt}
		\end{figure*}
		Combining the message from the Markov chain and the factor node $p(x_n|\alpha_n)$, the belief of $\alpha_{n}$ is denotes by (\ref{belief}), where
		\begin{equation}
			\psi_n^{\mathrm{post}} = \frac{\pi_n^{\mathrm{out}}\psi^f_n \psi^b_n}{\pi_n^{\mathrm{out}}\psi^f_n \psi^b_n + (1-\pi_n^{\mathrm{out}})(1-\psi^f_n)(1-\psi^b_n)},
			\label{lpost1}
		\end{equation}
		with $\psi_n^{\mathrm{post}}$ indicating that the estimates of the probability
		that the $n$-th antenna is visible. If $\psi_n^{\mathrm{post}}$ is greater than a threshold, the $n$-th antenna is regarded as being visible, i.e.,
		\begin{equation}
			\hat{{\alpha}}_n = \left\{
			\begin{aligned}
				1&,\quad \psi_n^{\mathrm{post}} > \psi_{\mathrm{th}}, \\
				0&,\quad \psi_n^{\mathrm{post}} \le \psi_{\mathrm{th}},
			\end{aligned}
			\right.
			\label{lpost2}
		\end{equation}
		where $\psi_{\mathrm{th}}$ is a predetermined threshold.
		
		In the following, we continue to formulate the backward passing, i.e., from right to left. According to the sum-product rule, the message from $\alpha_n$ to $p(x_n|\alpha_n)$ is given by
		\begin{equation}
			\begin{aligned}
				&\nu_{\alpha_n \rightarrow p(x_n|\alpha_n)}(\alpha_n) \\=& \nu_{p(\alpha_{n}|\alpha_{n-1}) \rightarrow \alpha_n}(\alpha_n) \nu_{p(\alpha_{n+1}|\alpha_{n}) \rightarrow \alpha_{n}}(\alpha_n) \\
				=&\psi^f_n \psi^b_n\delta(1-\alpha_{n})+(1-\psi^f_n)(1-\psi^b_n)\delta(\alpha_{n})\\
				\propto& \pi^{\mathrm{in}}_n \delta(1-\alpha_{n}) + (1-\pi^{\mathrm{in}}_n) \delta(\alpha_{n}),
			\end{aligned}
			\label{vap1}
		\end{equation}
		where $\pi^{\mathrm{in}}_n$ is defined as
		\begin{equation}
			\pi^{\mathrm{in}}_n = \frac{\psi^f_n \psi^b_n}{\psi^f_n \psi^b_n + (1-\psi^f_n)(1-\psi^b_n)}.
			\label{vap2}
		\end{equation}
		 Utilizing (\ref{vap1}) and the sum-product rule, the message from $p(x_n|\alpha_n)$ to $x_n$ is calculated as 
		\begin{equation}
			\begin{aligned}
				&\nu_{p(x_n|\alpha_n) \rightarrow x_n}(x_n)\\
				=& \int\nu_{\alpha_n \rightarrow p(x_n|\alpha_n)}(\alpha_n) p(x_n|\alpha_n) \mathrm{d}\alpha_{n}\\
				=&\pi^{\mathrm{in}}_n\mathcal{CN}(x_n; 0, \sigma^2) + (1-\pi^{\mathrm{in}}_n)\delta(x_n).
			\end{aligned}
			\label{vpx}
		\end{equation}
	
	\begin{algorithm}
		\renewcommand{\algorithmicrequire}{\textbf{Input:}}
		\renewcommand{\algorithmicensure}{\textbf{Output:}}
		\caption{VRDO-MP Algorithm}
		\begin{algorithmic}[1]
			\Require the received vector $\mathbf{y}$ and analog matrix $\mathbf{A}$; 
			\Statex \textbf{Initialize:} $\mathbf{x}_{A}^{\mathrm{pri}} \leftarrow \mathbf{0}$, $v_{A}^{\mathrm{pri}} \leftarrow 1$, and $\boldsymbol{\vartheta}$. 
			\While{the stopping criterion is not met}
			\Statex /*\textbf{Module A: LMMSE Estimator of $x_n$}*/
			\State Update $\mathbf{x}_{A}^{\mathrm{post}}$ and $v_{A}^{\mathrm{post}}$ by (\ref{xapost}) and (\ref{vapost});
			\State Update $\mathbf{x}_{A}^{\mathrm{ext}}$ and $v_{A}^{\mathrm{ext}}$ by (\ref{vaext});
			\State Update $\mathbf{x}_{B}^{\mathrm{pri}}$ and ${v}_{B}^{\mathrm{pri}}$ by $\mathbf{x}_{B}^{\mathrm{pri}} \leftarrow \mathbf{x}_{A}^{\mathrm{ext}}$ and ${v}_{B}^{\mathrm{pri}} \leftarrow {v}_{A}^{\mathrm{ext}}$;
			\Statex /*\textbf{Module B: Detector of} $\alpha_n$*/
			\State Update the message $\nu_{x_n \rightarrow p(x_n|\alpha_n)}(x_n)$ by (\ref{vxp});
			\State Update the message $\nu_{p(x_n|\alpha_n) \rightarrow \alpha_n}(\alpha_n)$ by (\ref{vpa1});
			\State Update the forward message passing of Markov chain by (\ref{Mf1}) and (\ref{Mf2});
			\State Update the backward message passing of Markov chain by (\ref{Mb1}) and (\ref{Mb2});
			\State Update $\nu_{\alpha_n \rightarrow p(x_n|\alpha_n)}$ by (\ref{vap1}) and (\ref{vap2});
			\State Update $\nu_{p(x_n|\alpha_n) \rightarrow x_n}$ by (\ref{vpx});
			\State Calculate $x_{n,B}^{\mathrm{post}}$ and $v_{B}^{\mathrm{post}}$ by (\ref{xBpost1}) and (\ref{vBpost2});
			\State Update $\mathbf{x}_{B}^{\mathrm{ext}}$ and $v_{B}^{\mathrm{ext}}$ by (\ref{vBext});
			\State Update $\mathbf{x}_{A}^{\mathrm{pri}}$ and ${v}_{A}^{\mathrm{pri}}$ by $\mathbf{x}_{A}^{\mathrm{pri}} \leftarrow \mathbf{x}_{B}^{\mathrm{ext}}$ and ${v}_{A}^{\mathrm{pri}} \leftarrow {v}_{B}^{\mathrm{ext}}$;
			\Statex /*\textbf{Parameters learning}*/
			\State Update hyperparameter $\boldsymbol{\vartheta}$ by the EM algorithm;
			\EndWhile
			\Statex /*\textbf{Estimation of} $\alpha_n$*/
			\State Calculate $\psi^n_{\mathrm{post}}$ and $\hat{\boldsymbol{\alpha}}$ by (\ref{lpost1}) and (\ref{lpost2}).
			\Ensure visibility probability ${\psi}_n^{\mathrm{post}}$ and indicator variable $\hat{{\alpha}}_n$
		\end{algorithmic}
		\label{Turbo}
	\end{algorithm}
	
	To realize the MP from Module B to Module A, we further elaborate the posteriori distribution of $\mathbf{x}$.
	Utilizing (\ref{vpx1}) and (\ref{vpx}), the posteriori mean $x_{n,B}^{\mathrm{post}} = \mathbb{E}\left\lbrace x_n|x_{n,B}^{\mathrm{pri}} \right\rbrace$ of $x_n$ can be given by 
	\begin{equation}
		\begin{aligned}
			x_{n,B}^{\mathrm{post}} = \mathbb{E}\left\lbrace x_n|x_{n,B}^{\mathrm{pri}} \right\rbrace
			= \frac{a_n x_{n,B}^{\mathrm{pri}}}{1+b_n q_n},
		\end{aligned}
		\label{xBpost1}
	\end{equation}
	where $a_n$, $b_n$, $c_n$ and $q_n$ are intermediate variables, satisfying  
	$a_n = {\sigma^2}/({v_{n,B}^{\mathrm{pri}}+\sigma^2})$, $b_n = {(1-\pi^{\mathrm{in}}_n)(v_{n,B}^{\mathrm{pri}}+\sigma^2)}/{\pi^{\mathrm{in}}_n v_{n,B}^{\mathrm{pri}}}$, $c_n = {\sigma^2}/{v_{n,B}^{\mathrm{pri}}(\sigma^2+v_{n,B}^{\mathrm{pri}})}$, and $q_n = \exp (-c_n| x_{n,B}^{\mathrm{pri}}|^2)$.

	Similarly, the posterior variance $v_{n,B}^{\mathrm{post}} = \mathbb{V}\mathrm{ar}\left\lbrace x_n|x_{n,B}^{\mathrm{pri}} \right\rbrace$ of $x_n$ are given by
	\begin{align}
		\label{vBpost1} v_{n,B}^{\mathrm{post}} &= \mathbb{V}\mathrm{ar}\left\lbrace x_n|x_{n,B}^{\mathrm{pri}} \right\rbrace = b_n q_n \left|x_{n,B}^{\mathrm{post}} \right|^2 + \frac{v_{n,B}^{\mathrm{pri}}}{x_{n,B}^{\mathrm{pri}}}x_{n,B}^{\mathrm{post}},\\
		\label{vBpost2} v_{B}^{\mathrm{post}} &= \frac{1}{N} \sum_{n =1}^N v_{n,B}^{\mathrm{post}}.
	\end{align}
	With the above posterior mean and variance, the extrinsic outputs
	of Module B can be obtained as follows
	\begin{align}
		v_{B}^{\mathrm{ext}} = \label{vBext} \frac{v_{B}^{\mathrm{pri}}v_{B}^{\mathrm{post}}}{v_{B}^{\mathrm{pri}}-v_{B}^{\mathrm{post}}}, \,
		\mathbf{x}_{B}^{\mathrm{ext}} = v_{B}^{\mathrm{ext}}\left(\frac{\mathbf{x}_{B}^{\mathrm{post}}}{v_{B}^{\mathrm{post}}} -\frac {\mathbf{x}_{B}^{\mathrm{post}}}{v_{B}^{\mathrm{pri}}}\right).
	\end{align}
	
Additionally, the VRDO-MP algorithm requires the prior knowledge of the system statistics characterized by the hyperparameter set $\boldsymbol{\vartheta} = \{\sigma^2, \sigma_{\mathrm{N}}^2, \psi ,p_{10}\}$, where $\sigma^2$ is the variance of spatial-domain channels, $\sigma_{\mathrm{N}}^2$ is the noise variance, $\psi$ is sparse level of spatial-domain channels and $p_{01}$ is the state transition probability from $\alpha_n = 0$ to $\alpha_n = 1$. To obtain these prior knowledge, we propose to integrate the EM algorithm \cite{EM_Algorithm} into the VRDO-MP algorithm, where the statistical parameters are learned from the received signals during the estimation procedure. In summary, the proposed VRDO-MP algorithm can be organized in a more succinct form, which is summarized in Algorithm \ref{Turbo} and can be terminated when it reaches a maximum number of iteration or the difference between the estimates of two consecutive iterations is less than a threshold.

\begin{remark}
	While both this work and \cite{HMMLiuAn2} utilize the Turbo based MP scheme for MAP estimation, it is crucial to highlight the fundamental differences in problem derivation between the two studies. In \cite{HMMLiuAn2}, the Turbo based MP scheme is employed to estimate angular-domain channel coefficients. However, in this work, the focus of the Turbo based MP scheme is on antenna visibility detection. Additionally, compared to \cite{HMMLiuAn2}, the proposed VRDO-MP algorithm further derives the belief of $\alpha_n$. It is important to emphasize that the derivation of the belief in $\alpha_n$ is a novel contribution and has not been previously reported in the literature.
\end{remark}
\vspace{-1em}
\subsection{Stage \uppercase\expandafter{\romannumeral2}: CE in Wavenumber Domain}
\begin{algorithm}
	\renewcommand{\algorithmicrequire}{\textbf{Input:}}
	\renewcommand{\algorithmicensure}{\textbf{Output:}}
	\caption{BB-OMP Algorithm}
	\begin{algorithmic}
		\Require received vector $\mathbf{y}$ and analog combiner matrix $\mathbf{A}$, wavenumber-domain codebook $\mathbf{F}$, and the visibility probability $\boldsymbol{\psi}^{\mathrm{post}}$;
		\Statex \textbf{Initialize:} $\mathbf{r}_{-1} = \mathbf{0}$, $\mathbf{r}_{0} = \mathbf{y}$, $\mathcal{I}_0 = \emptyset$, and $t=1$;
		\State Construct the sensing matrix $\hat{\boldsymbol{\Phi}}$ by (\ref{Phi});
		\While{the stopping criterion is not met}
		\State $j = \mathop{\arg\min}\limits_{i} \left|\hat{\boldsymbol{\Phi}}(:,i)^{\mathrm{H}} \mathbf{r}_{t-1} \right|$, $i \in \left\lbrace1,2,\cdots, S\right\rbrace $;
		\State $\mathcal{I}_t = \mathcal{I}_{t-1} \cup j$;
		\State $\mathbf{h}_{a,t}^{\mathrm{SnS}} = \mathop{\arg\min}\limits_{\mathbf{h}_{a,t}^{\mathrm{SnS}}} \lVert \mathbf{y}-\hat{\boldsymbol{\Phi}}_{\mathcal{I}_t}\mathbf{h}_{a,t}^{\mathrm{SnS}} \rVert_2$;
		\State $\mathbf{r}_t = \mathbf{y} - \hat{\boldsymbol{\Phi}}_{\mathcal{I}_t} \mathbf{h}_{a,t}^{\mathrm{SnS}}$;
		\State $t = t+1$;
		\EndWhile
		\State $\hat{\mathbf{h}}_a^{\mathrm{SnS}}(i) = \mathbf{h}_{a,t-1}^{\mathrm{SnS}}(i)$ for $i \in \mathcal{I}_{t-1}$ and $\hat{\mathbf{h}}_a^{\mathrm{SnS}}(i) =0$ otherwise;
		\State $\hat{\mathbf{x}} = \mathrm{diag}(\boldsymbol{\psi}_{\mathrm{post}})\mathbf{F}\hat{\mathbf{h}}_a^{\mathrm{SnS}}$.
		\Ensure $\hat{\mathbf{x}}$
	\end{algorithmic}
	\label{CS_OMP}
\end{algorithm}
	In the second stage, we attempt to estimate $\mathbf{h}_a^{\mathrm{SnS}}$ by exploiting the VR information from the first stage and the wavenumber-domain sparsity. Specifically, in the second stage, we have
	\begin{equation}
		\mathbf{y}  =\mathbf{A}\mathrm{diag}(\hat{\boldsymbol{\alpha}})\mathbf{F}\mathbf{h}_a^{\mathrm{SnS}} + \mathbf{n} = \boldsymbol{\Phi}\mathbf{h}_a^{\mathrm{SnS}} + \mathbf{n},
		\label{pilot4}
	\end{equation}
	where $\hat{\boldsymbol{\alpha}} = [\hat{{\alpha}}_1, \hat{{\alpha}}_2, \cdots, \hat{{\alpha}}_N]^{\mathrm{T}}$ and $\boldsymbol{\Phi} = \mathbf{A}\mathrm{diag}(\hat{\boldsymbol{\alpha}})\mathbf{F}$ denotes the equivalent measurement matrix. Since $\mathbf{h}_a^{\mathrm{SnS}}$ is sparse in the wavenumber domain, the estimation problem of $\mathbf{h}_a^{\mathrm{SnS}}$ from (\ref{pilot4}) can be also formulated and then solved as a sparse signal recovery problem, i.e.,
	\begin{equation}
		\begin{aligned}
			\hat{\mathbf{h}}_a^{\mathrm{SnS}} =\mathop{\arg\min}\limits_{\mathbf{h}_a^{\mathrm{SnS}}} & \quad \lVert \mathbf{y}-{\boldsymbol{\Phi}}\mathbf{h}_a^{\mathrm{SnS}} \rVert_2 \\
			\mathrm{s.t.} & \quad \lVert\mathbf{h}_a^{\mathrm{SnS}} \rVert_0 = L_e^{\mathrm{SnS}},
		\end{aligned}
		\label{OMP}
	\end{equation} 
	where $L_e^{\mathrm{SnS}}$ is the spatial level of $\mathbf{h}_a^{\mathrm{SnS}}$ in wavenumber domain. Inspired by the soft decision in channel coding and decoding theory \cite{MP_Contributions}, instead of utilizing $\hat{\boldsymbol{\alpha}}$ directly, we try to use the soft information of probability  $\psi_n^{\mathrm{post}}$ to reconstruct the equivalent measurement matrix $\hat{\boldsymbol{\Phi}}$ in (\ref{pilot4}), i.e., 
	\begin{equation}
		\hat{\boldsymbol{\Phi}} = \mathbf{A}\mathrm{diag}(\boldsymbol{\psi}^\mathrm{post})\mathbf{F},
		\label{Phi}
	\end{equation}  
	where $\boldsymbol{\psi}^\mathrm{post} = [\psi_1^{\mathrm{post}}, \psi_2^{\mathrm{post}},\cdots,\psi_N^{\mathrm{post}}]^{\mathrm{T}}$ indicates the belief of antenna visibility.

	The optimization problem in (\ref{OMP}) is a non-convex optimization with $L_0$ norm and is difficult to solve. To address this, we propose the BB-OMP algorithm to solve (\ref{OMP}). The details are summarized in Algorithm~\ref{CS_OMP}, where the termination criterion is similar to  Algorithm~\ref{Turbo}. Compared to the traditional OMP method \cite{PolarCS}, the proposed BB-OMP algorithm is characterized by the belief-based measurement matrix, which implies the VR information of the antenna domain. Therefore, the estimation performance can be greatly improved, as will be verified by simulations in Section \ref{section5}.
	\vspace{-1em}
	\subsection{Overall Algorithm Description}
	\begin{figure}
		\centering
		\includegraphics[width=0.5\textwidth]{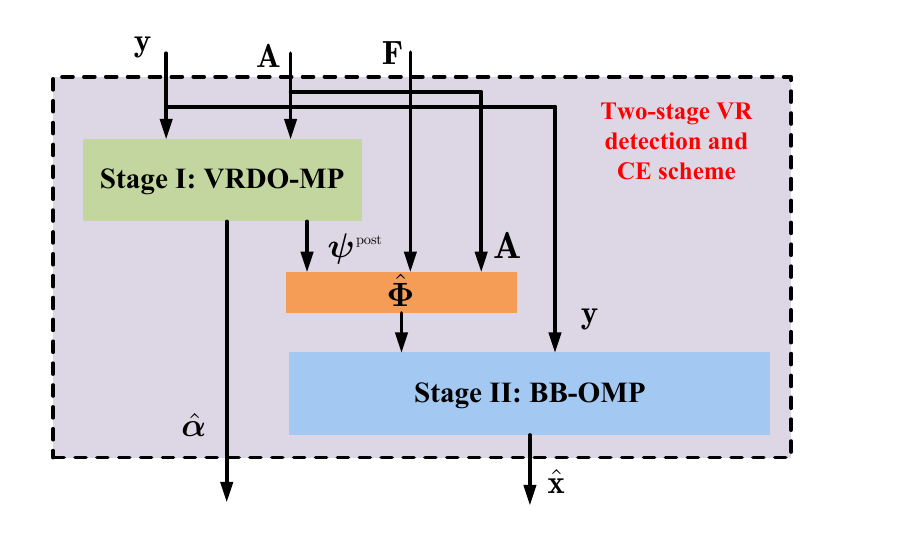}
		\caption{Illustration of the proposed two-stage VR detection and CE algorithm.}
		\label{Algorithm_Flow}
	\end{figure}
	\begin{algorithm}
		\renewcommand{\algorithmicrequire}{\textbf{Input:}}
		\renewcommand{\algorithmicensure}{\textbf{Output:}}
		\caption{TS-VDCE Algorithm}
		\begin{algorithmic}
			\State \textbf{STAGE 1:} Operating in the \textbf{antenna domain}
			\Statex \begin{enumerate}
				\item input: received vector $\mathbf{y}$ and {analog combiner matrix} $\mathbf{A}$;
				\item obtain the visibility probability $\boldsymbol{\psi}^{\mathrm{post}}$ and VR $\hat{\boldsymbol{\alpha}}$ using Algorithm \ref{Turbo};
			\end{enumerate}
			\State \textbf{STAGE 2:} Operating in \textbf{wavenumber domain}
			\Statex \begin{enumerate}
				\item input: the received vector $\mathbf{y}$, {analog combiner matrix} $\mathbf{A}$, wavenumber-domain codebook $\mathbf{F}$, and the visibility probability $\boldsymbol{\psi}^{\mathrm{post}}$;
				\item obtain the channel estimate $\hat{\mathbf{x}}$ using Algorithm \ref{CS_OMP};
			\end{enumerate}
			\State \textbf{Outputs:}
			\Statex \begin{enumerate}
				\item the VR indicator vector $\hat{\boldsymbol{\alpha}}$;
				\item the channel estimate $\hat{\mathbf{x}}$.
			\end{enumerate}
		\end{algorithmic}
		\label{TS}
	\end{algorithm}
	The TS-VDCE scheme is illustrated in Fig. \ref{Algorithm_Flow}. Specifically, the VR information $\hat{\boldsymbol{\alpha}}$ and $\boldsymbol{\psi}^{\mathrm{post}}$ are first obtained through VRDO-MP algorithm according to the observation vector $\mathbf{y}$ and analog combiner matrix $\mathbf{A}$. Subsequently, based on the obtained VR information $\boldsymbol{\psi}^{\mathrm{post}}$,  the equivalent measurement matrix $\hat{\boldsymbol{\Phi}}$ can be obtained. Finally, the wavenumber-domain channel and corresponding spatial-domain channel can be estimated through the BB-OMP algorithm. 
	The detailed steps are summarized in Algorithm \ref{TS}.
	In the following, we provide the complexity analysis for the proposed TS-VRCE algorithm. Under the setting that $\mathbf{y} \in \mathbb{C}^{QN_{\mathrm{rf}}\times1}$, $\mathbf{A} \in \mathbb{C}^{QN_{\mathrm{rf}} \times N}$, $\mathbf{F} \in \mathbb{C}^{N\times J}$, and $\mathbf{x}\in \mathbb{C}^{N\times 1}$, the complexity is elaborated as follows. In the stage I, the computational complexity for Module A per iteration is $\mathcal{O}(QN_{\mathrm{rf}}N)$. For module B, the
	complexity of the MP per iteration is
	$\mathcal{O}(N \log N)$. Therefore, the overall complexity of Algorithm \ref{Turbo} is $\mathcal{O}(I_1QN_{\mathrm{rf}}N)$, where $I_1$
	denotes the number of iterations in stage I. In the stage II, the computational complexity of Algorithm \ref{CS_OMP} per iteration is $\mathcal{O}(QN_{\mathrm{rf}}J)$. Thus, the overall complexity of Algorithm \ref{TS} is $\mathcal{O}(I_1QN_{\mathrm{rf}}N + I_2 QN_{\mathrm{rf}}J)$, where $I_2$
	denotes the number of iterations in stage~II.
\section{Simulation Results}
\label{section5}
In this section, we evaluate the performance of the proposed two-stage VR detection and CE scheme under various system setups. In particular, we consider the VR error ratio (VRER) and normalized mean square error (NMSE) as performance metrics to evaluate the VR detection and CE performance, respectively. The VRER and NMSE are defined respectively as $\mathrm{VRER} \triangleq {\sum_{n =1}^N \left|\alpha_n - \hat{{\alpha}}_n\right|}/{N} \le 1$ and $\mathrm{NMSE} \triangleq {\lVert \hat{\mathbf{x}} - \mathbf{x} \rVert^2_2}/{\lVert \mathbf{x} \lVert^2_2}$,
where $\hat{\boldsymbol{\alpha}}$ and $\hat{\mathbf{x}}$ denote the VR indicator vector and estimated channel, respectively;  $\boldsymbol{\alpha}$ and ${\mathbf{x}}$ denote the true VR indicator vector and true channel vector.
	
The following parameters are utilized throughout the section unless specified otherwise: the number of antenna elements $N=256$, the number of RF chains $N_{\mathrm{rf}} = 4$, the carrier frequency $f_c = 100$GHz, user number $P=4$, and the oversampling factor $S=2$. We assume that users are randomly distributed in a sector region of BS with a direction angle from $-\pi/3$ to $\pi/3$ and a distance from 10m to 50m. All the numerical results here are obtained by averaging over 5000 channel realizations.
\vspace{-1em}
\subsection{VR Detection Performance}
To verify the effectiveness of the developed VR detection algorithm, we utilize the rising and falling edges-based (RFEB) baseline proposed in \cite{HanYu}. Note that the RFEB method is based on the CE result of the MJCE algorithm in \cite{MJCE}. However, the MJCE method cannot be applied to hybrid precoding architectures. Therefore, for comparison, we utilize the results of the least squares (LS) estimator to evaluate the performance of the RFEB method here.
Fig. \ref{VRER1} illustrates the VRER performance against SNR. It is evident that the proposed VRDO-MP algorithm exhibits the pronounced superiority compared with the baseline algorithm.
This reason is that the RFEB method relies solely on the statistical characteristics of the received power, making it sensitive to noise levels. 
Specifically, in low SNR scenarios, the RFEB method is prone to obtaining fake rising or falling edges. 
In contrast, the VRDO-MP algorithm exploits the spatial correlation of VR through the Markov chain. 
Consequently, it proves to be more robust to varying noise levels. 
For example, with $\mathrm{SNR} = 0$dB,  $Q = 35$ and $\psi = 0.25$, the average correct detection ratio of VRDO-MP exceeds 90\%, while RFEB achieves only around 70\%, as depicted in Fig. \ref{VRER1a}. 
\begin{figure}
	\centering
	\subfigure[$\psi =0.125$]{
		\includegraphics[width=0.22\textwidth]{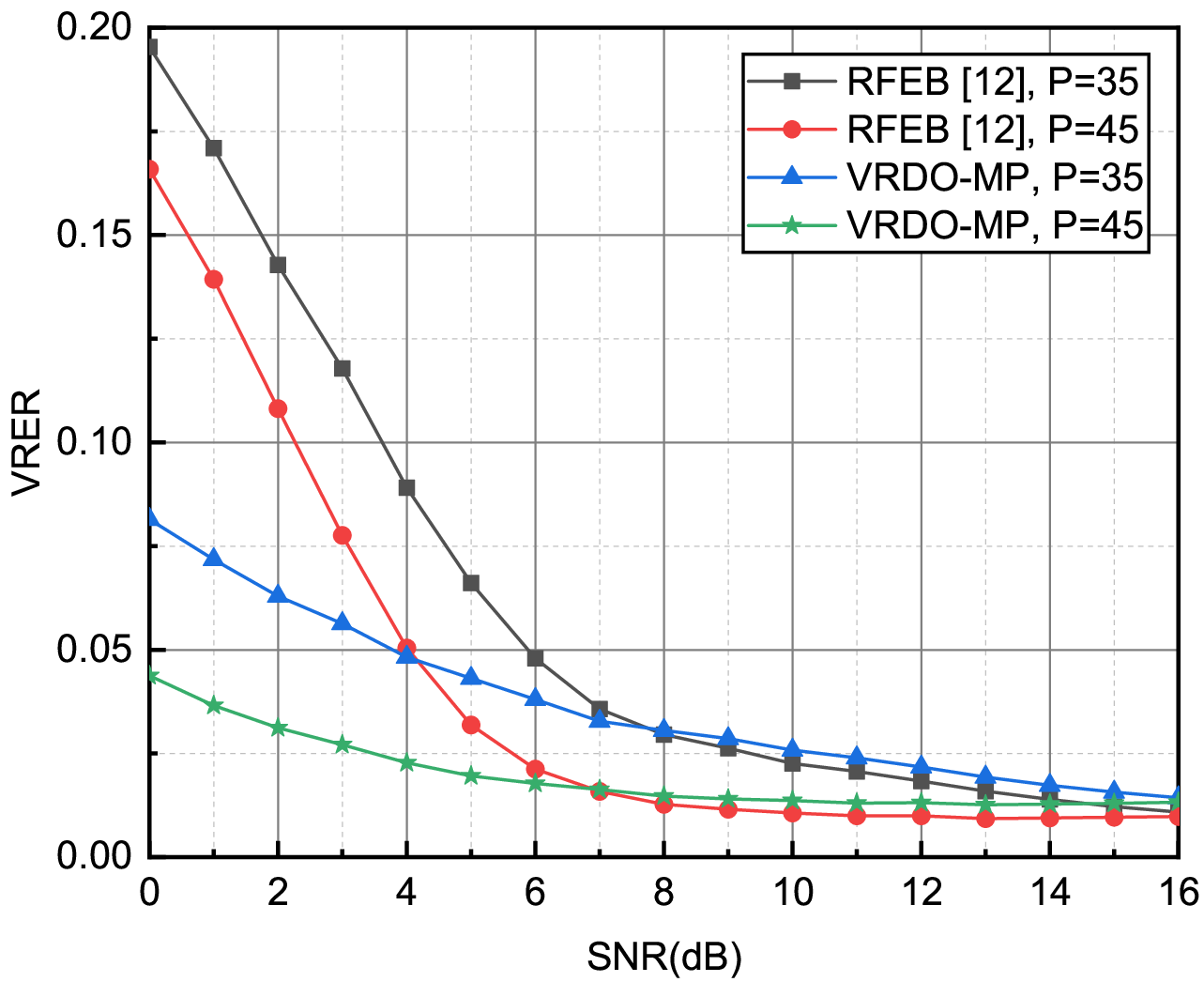}
		\label{VRER1b}
	}
	\subfigure[$\psi = 0.25$]{
		\includegraphics[width=0.22\textwidth]{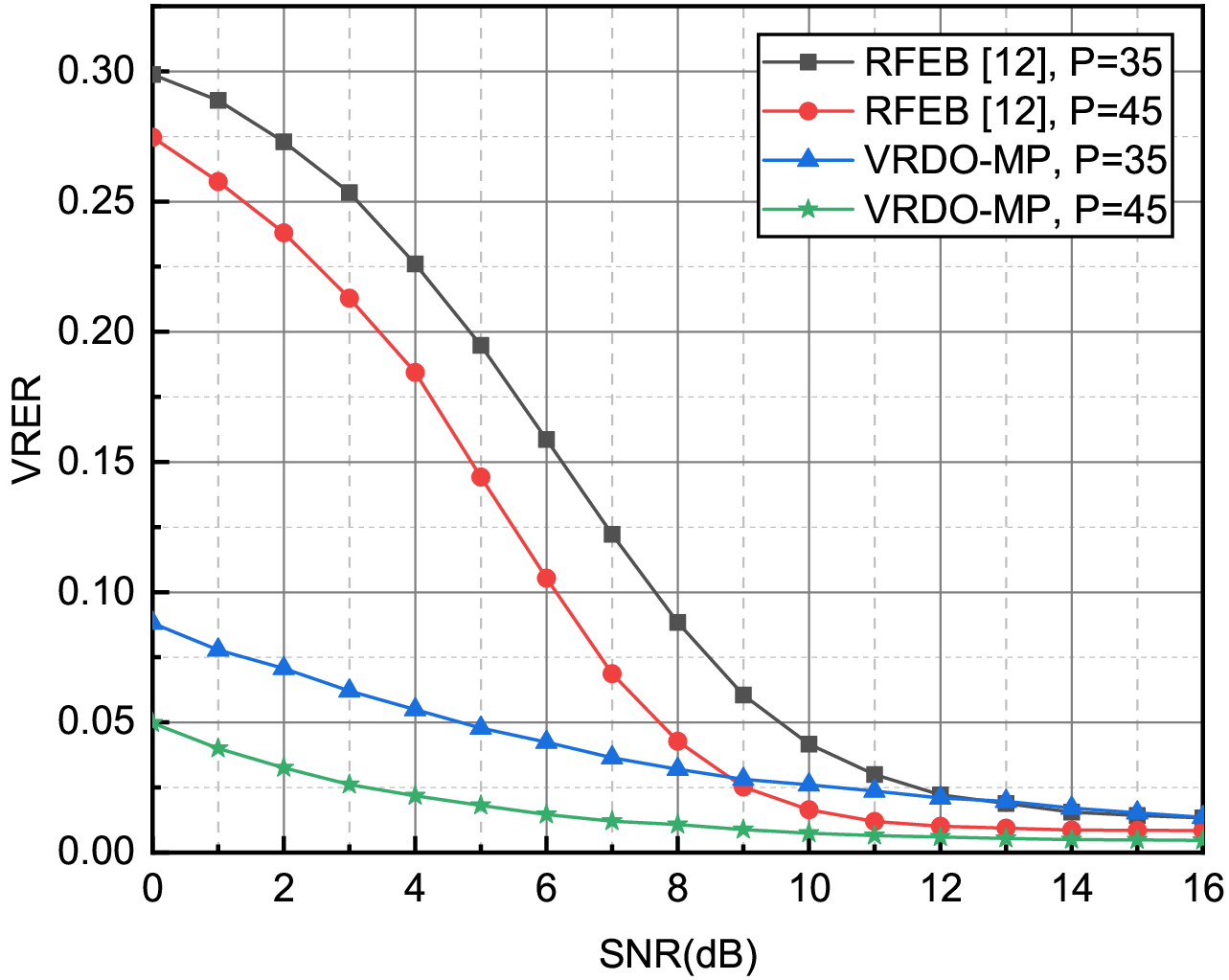}
		\label{VRER1a}
	}
	\caption{VRER performance versus SNR}
	\label{VRER1}
\end{figure}

\begin{figure}
	\centering
	\subfigure[Continuous VR]{
		\includegraphics[width=0.22\textwidth]{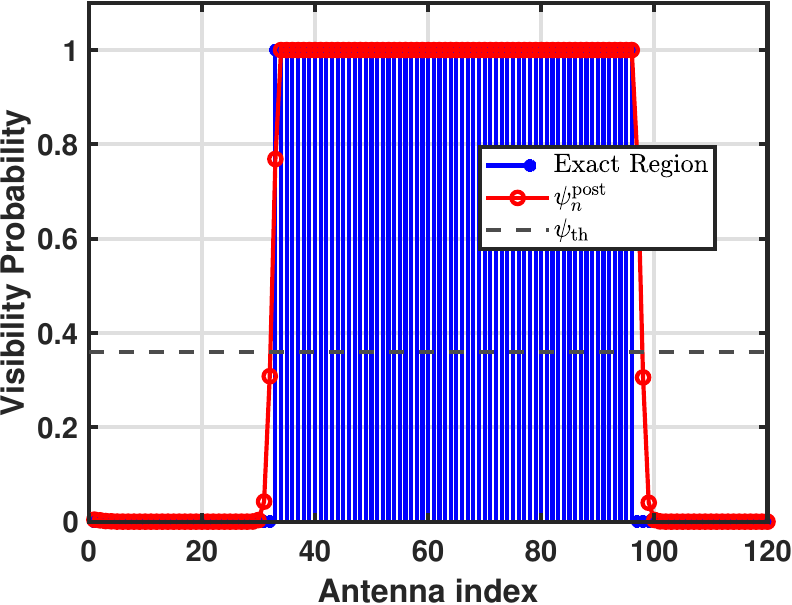}
		\label{VRER2a}
	}
	\subfigure[Discontinuous  VR]{
		\includegraphics[width=0.22\textwidth]{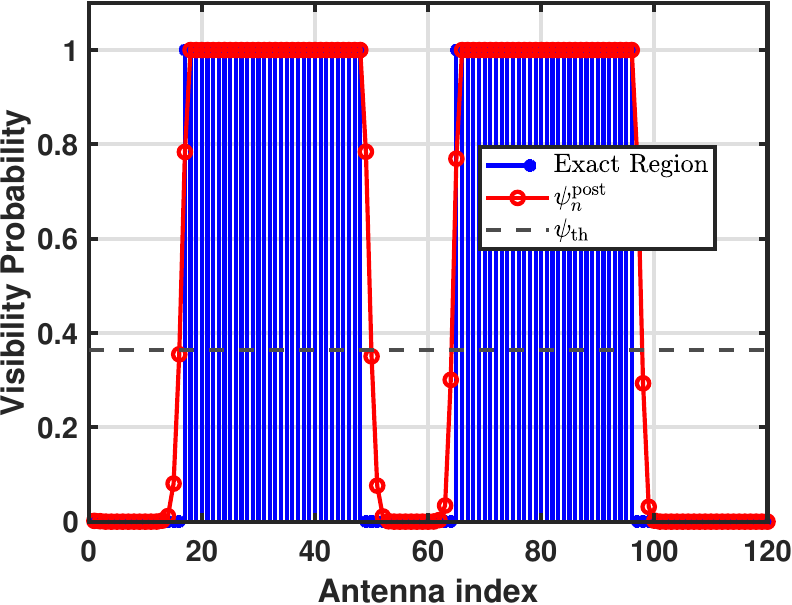}
		\label{VRER2b}
	}
	\caption{Visibility probability of antenna elements}
	\label{VRER2}
\end{figure}

Furthermore, Fig. \ref{VRER2} presents the results of a Monte Carlo simulation with parameters set to $\mathrm{SNR}=5$ dB, $Q=45$, and $\psi = 0.25$. For the case of continuous VR, the set of visible antenna element is set to $\boldsymbol{\phi}_{\mathrm{c}} = \left\lbrace33,34,\cdots,95,96 \right\rbrace$. In terms of discontinuous VRs, the set of visible antenna element is set to $\boldsymbol{\phi}_{\mathrm{d}} = \boldsymbol{\phi}^1 \cup \boldsymbol{\phi}^2$ with $\boldsymbol{\phi}^1 = \left\lbrace17,18,\cdots,47,48 \right\rbrace$ and $\boldsymbol{\phi}^2 = \left\lbrace65,66,\cdots,95,96 \right\rbrace$.
From Fig. \ref{VRER2a}, it is observed that the detection probability $\psi^{\mathrm{post}}_n$ is evidently close to 1 within the VR and close to 0 outside the VR, which confirms the efficacy of the proposed detection algorithm. Additionally, the VRDO-MP algorithm also proves to be feasible for discontinuous VRs, as illustrated in Fig. \ref{VRER2b}.
This capability enables the algorithm to seamlessly handle multipath scenarios, where different scatterers correspond to different VRs.
\vspace{-1em}
\subsection{CE Performance}
\begin{table}
	\renewcommand{\arraystretch}{1.5}
	\centering
	\caption{Computational Complexity Comparison}
	\begin{tabular}{|c|c|}
		\hline
		Algorithms         & Computational complexity $\mathcal{O}(\cdot)$ \\ \hline
		LS                 & $\mathcal{O}(NQ^2 N^2_{\mathrm{rf}}+Q^3 N^3_{\mathrm{rf}}+Q N_{\mathrm{rf}}N)$                        \\ \hline
		P-OMP              & $\mathcal{O}(I_p Q N_{\mathrm{rf}} J_\mathrm{polar})$                        \\ \hline
		W-OMP              & $\mathcal{O}(I_w Q N_{\mathrm{rf}} J_\mathrm{wave})$                        \\ \hline
		AD-AMP              & $\mathcal{O}(I_a Q N_{\mathrm{rf}} N)$                        \\ \hline
		Genie-aided W-OMP & $\mathcal{O}(I_g Q N_{\mathrm{rf}} J_\mathrm{wave})$                        \\ \hline
		Proposed TS-VDCE        & $\mathcal{O}(I_1Q N_{\mathrm{rf}}N + I_2 Q N_{\mathrm{rf}}J_{\mathrm{wave}})$                       \\ \hline
	\end{tabular}
	\label{Comparsion}
\end{table}

To the evaluate CE performance, we compare our proposed strategy with the following benchmarks. 
\begin{itemize}
	\item \textbf{LS}: Least square estimator based on the formulation (\ref{pilot3}).
	\item \textbf{P-OMP}: On-grid polar-domain simultaneous orthogonal matching pursuit algorithm for XL-MIMO channel proposed in \cite{PolarCS} without considering VR.
	\item \textbf{W-OMP}: On-grid wavenumber-domain simultaneous orthogonal matching pursuit algorithm. Similar to P-OMP, only the polar-domain codebook is replaced by wavenumber-domain codebook.
	
	\item \textbf{AD-AMP}: Antenna-domain AMP algorithm proposed in~\cite{Bayesian1}. Compared with AD-AMP, the proposed algorithm operates simultaneously in the antenna and wavenumber domains to fully capture the structured sparsity of SnS XL-MIMO channels.
	\item \textbf{Genie-aided W-OMP}. W-OMP algorithm with perfect knowledge of VR as an absolute performance lower bound.
\end{itemize}

We commence our analysis by comparing the computational complexities of various estimation algorithms with $\mathbf{y} \in \mathbb{C}^{QN_{\mathrm{rf}}\times1}$, $\mathbf{A} \in \mathbb{C}^{QN_{\mathrm{rf}} \times N}$, $\mathbf{F}{\mathrm{p}} \in \mathbb{C}^{N\times J{\mathrm{p}}}$, $\mathbf{F}{\mathrm{w}} \in \mathbb{C}^{N\times J{\mathrm{w}}}$, and $\mathbf{x}\in \mathbb{C}^{N\times 1}$. Here, $\mathbf{F}{\mathrm{p}}$ and $\mathbf{F}{\mathrm{w}}$ represent polar-domain and wavenumber-domain codebooks, respectively, with $J_{\mathrm{p}} \gg J_{\mathrm{w}}$. The computational complexities are detailed in Table~\ref{Comparsion}, where $I_p$, $I_w$, $I_a$, $I_g$, $I_1$, and $I_2$ denote the number of iterations for each algorithm.
The LS algorithm, due to its matrix inversion operations, exhibits notably high computational complexity. Conversely, the remaining algorithms demonstrate comparable complexity, scaling linearly with $Q$, $N_{\mathrm{rf}}$, and $N$. It is imperative to note that despite a considerable increase in computational complexity, the proposed TS-VDCE algorithm manifests superior estimation performance, which will be verified later.

\begin{figure*}
	\centering
	\subfigure[NMSE  versus SNR]{
		\includegraphics[width=0.22\textwidth]{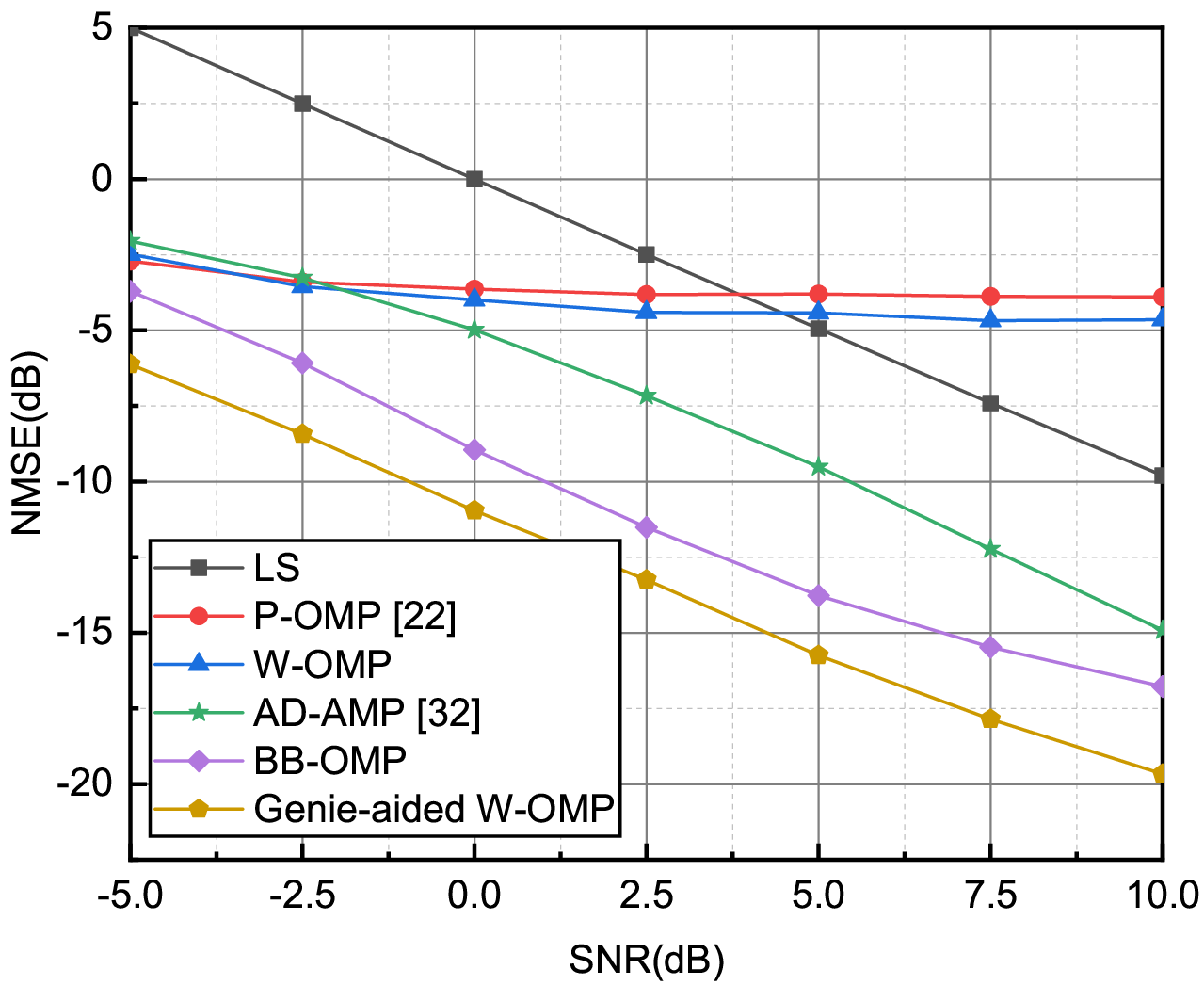}
		\label{NMSE1}
	}
	\subfigure[SE versus SNR]{
		\includegraphics[width=0.235\textwidth]{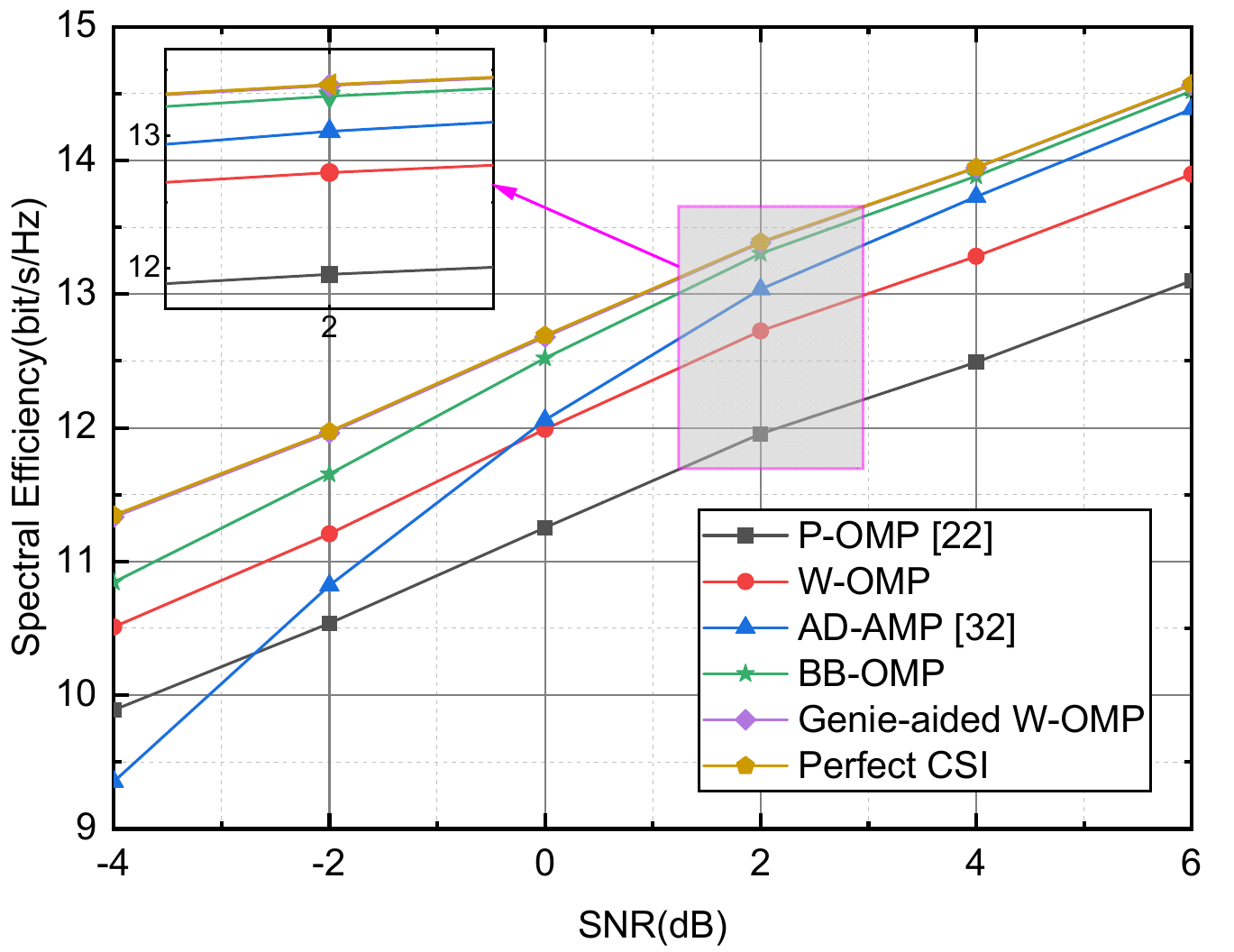}
		\label{capacity}
	}
	\subfigure[NMSE versus pilot length]{
		\includegraphics[width=0.22\textwidth]{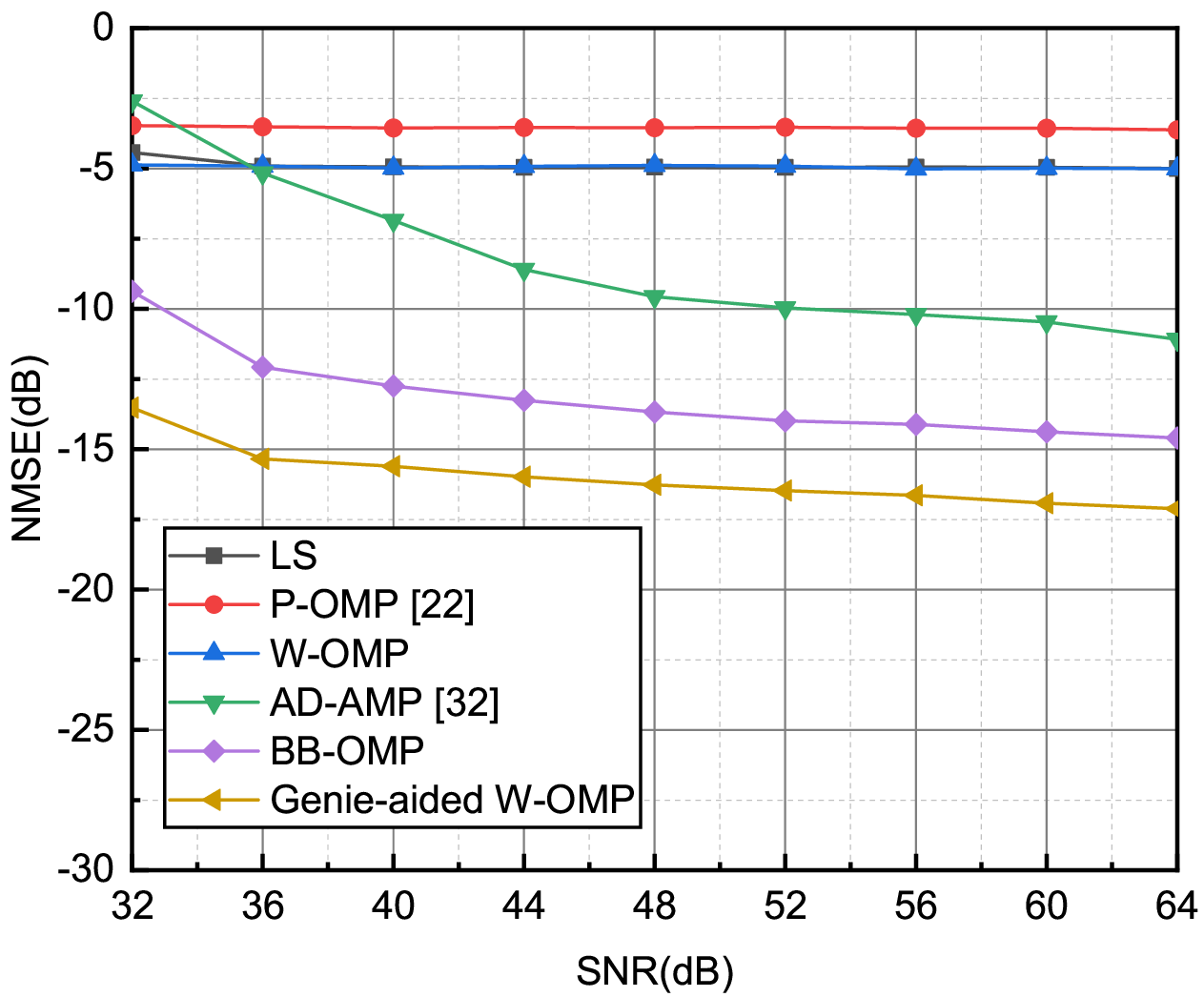}
		\label{NMSE2}
	}
	\subfigure[NMSE versus VR size]{
		\includegraphics[width=0.22\textwidth]{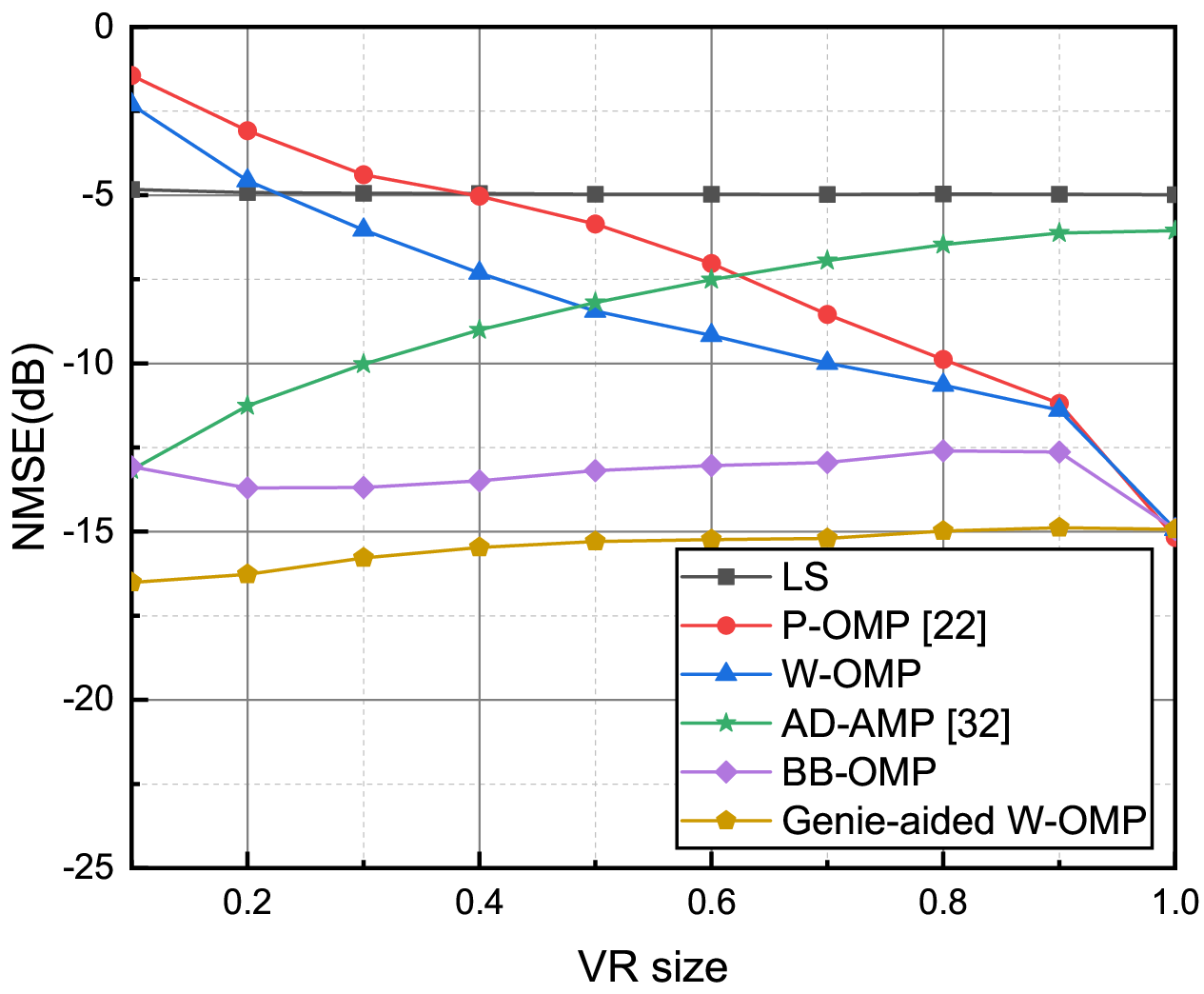}
		\label{NMSE_vr}
	}
	\label{NMSE}
\caption{NMSE versus different estimation algorithms}
\end{figure*}

Fig. \ref{NMSE1} illustrates NMSE performance against SNR with $Q=45$ and $\psi = 0.25$. The results reveal that channel estimation schemes neglecting the influence of SnS property experience severe performance degradation. Therefore, accurate VR information stands as a crucial prerequisite for precise channel estimation.
In addition, the proposed BB-OMP algorithm outperforms the existing spatial-domain algorithm and approaches the performance bound achieved by the Genie-aided W-OMP scheme. This notable improvement arises from the fact that, in contrast to the antenna-domain approach proposed in \cite{Bayesian1}, which relies solely on the sparsity of the antenna domain, the proposed estimation scheme adeptly leverages both antenna-domain spatial correlation and wavenumber-domain sparsity.

{To more intuitively demonstrate the performance superiority, we consider another metric: spectral efficiency (SE). For simplicity, we adopt the maximum ratio transmission (MRT) scheme for the downlink user data transmission\footnote{{Assume the analog precoding and digital precoding of the BS are represented by $\mathbf{F}_{\mathrm{RF}}\in \mathbb{C}^{N \times N_{\mathrm{RF}}}$ and $\mathbf{f}_{\mathrm{BB}} \in \mathbb{C}^{N{\mathrm{RF}}\times 1}$, respectively. According to the MRT scheme, we have $\mathbf{F}_{\mathrm{RF}}\mathbf{f}_{\mathrm{BB}} = \hat{\mathbf{x}}$.}}. 
Therefore, the spectral efficiency is given by $R=\log_2(1+\left|\mathbf{x}^{\mathrm{H}}\hat{\mathbf{x}}\right|^2/\sigma^2_{\mathrm{N}})$ \cite{MRT}.
Fig. \ref{capacity} shows the achievable SE versus SNR for different channel estimation algorithms when realistic channel realizations are considered. The simulation parameters are the same as those in Fig. \ref{NMSE1}. 
It is evident that the SE achieved by the proposed BB-OMP closely aligns with that achieved by Genie-aided W-OMP and perfect CSI. However, a noticeable performance gap exists between perfect CSI and P-OMP and W-OMP algorithms. This disparity arises because P-OMP and W-OMP algorithms do not consider VR detection in advance.
Additionally, while the SE performance of AD-AMP approaches that of BB-OMP as SNR increases, it is noted that AD-AMP is sensitive to the noise level due to its lack of consideration for wavenumber-domain sparsity. Consequently, at low SNR conditions, such as below 0 dB, the performance of AD-AMP significantly deteriorates.}
  
The performance comparison of the proposed algorithms
and the benchmarks at different pilot length $Q$ is evaluated in Fig. \ref{NMSE2} with $\mathrm{SNR} = 5$dB and $\psi = 0.25$.  The pilot sequence length $Q$ increases from $32$ to $64$, so the compressive ratio $QN_{\mathrm{rf}}/N$ correspondingly increases from $0.5$ to $1$. Similarly, we can see that the proposed BB-OMP algorithms significantly outperform the existing algorithms, especially when the pilot length is small, such as $Q=36$ or $Q=40$. This indicates that the proposed algorithms can significantly reduce the estimation overhead.

To quantify the effects of spatial non-stationarity, Fig.~\ref{NMSE_vr} further illustrates the NMSE performance versus VR size $\psi$ with $\mathrm{SNR} = 5$dB and $Q=45$. Notably, when $\psi = 1$, it indicates that the channel is spatially stationary. From Fig.~\ref{NMSE_vr}, it is apparent that, with increasing $\psi$, the performance of the AD-AMP algorithm deteriorates. This decline is attributed to the dependency of AD-AMP algorithm on the sparsity of the antenna domain. As $\psi$ increases, the sparsity of the antenna domain diminishes, leading to performance degradation.
In contrast, the performance of the proposed TS-VDCE algorithm exhibits only a slight fluctuation across varying $\psi$. {In other words, the TS-VDCE algorithm shows robustness to changes in the size of the VR. Thus, it adapts effectively to spatially stationary and SnS scenarios. Consequently, the TS-VRCE scheme is applicable to the near-field spatial stationary and spatially non-stationary scenarios. In addition, due to the robustness of TS-VRCE algorithm to VR size and the sparsity of far-field channels in wavenumber domain, the TS-VRCE algorithm is also applicable for the far-field channels in the LMMSE estimator.}
Furthermore, although the performance of the P-OMP and W-OMP algorithms improves with increasing $\psi$, they consistently lag behind the proposed algorithm due to the lack of VR information, which further validates the advantages of the two-stage estimation scheme.

\begin{figure}
	\centering
	\subfigure[$\psi = 0.125$]{
		\includegraphics[width=0.22\textwidth]{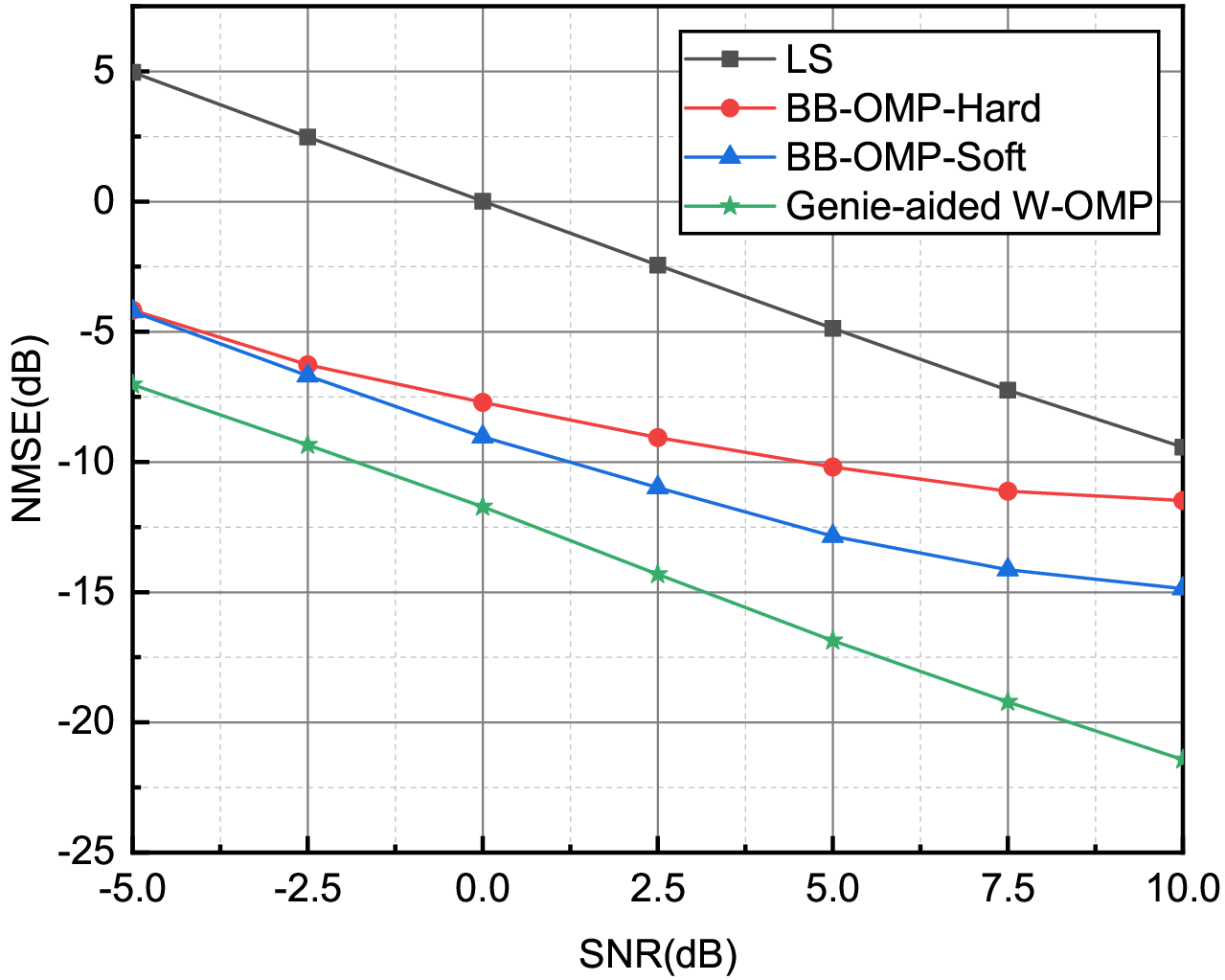}
		\label{NMSE3a}
	}
	\subfigure[$\psi = 0.25$]{
		\includegraphics[width=0.22\textwidth]{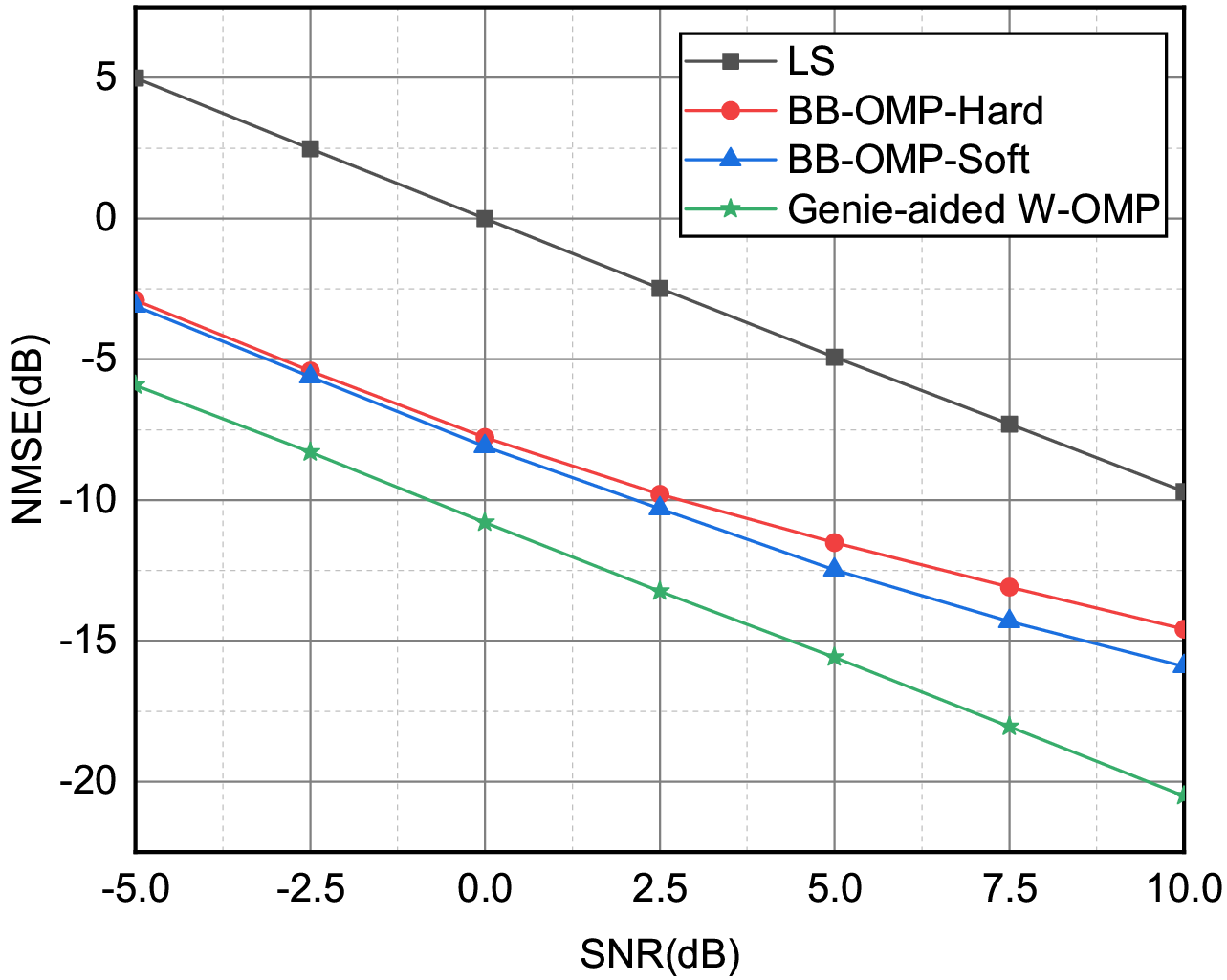}
		\label{NMSE3b}
	}
	\caption{NMSE versus different measurement matrices}
	\label{NMSE3}
\end{figure}
Fig. \ref{NMSE3} investigates the estimation performance against different measurement matrices with $Q=45$. Specifically, the measurement matrices are constructed by the following methods: 1) $\hat{\boldsymbol{\Phi}} = \mathbf{A}\mathrm{diag}(\boldsymbol{\psi}^\mathrm{post})\mathbf{F}$; 2) $\hat{\boldsymbol{\Phi}} = \mathbf{A}\mathrm{diag}(\hat{\boldsymbol{\alpha}})\mathbf{F}$, which correspond to the BB-OMP-Soft and BB-OMP-Hard algorithms, respectively. It can be seen that the BB-SOMP-Soft always outperforms BB-SOMP-Hard algorithms due to the soft decision is more robust for the noise level. Additionally, the more sparse the channel in the antenna domain, the more obvious the performance improvement, as shown in Fig.\ref{NMSE3a} and Fig.~\ref{NMSE3b}.  
\begin{figure}
	\centering
	\subfigure[$Q = 35$]{
		\includegraphics[width=0.22\textwidth]{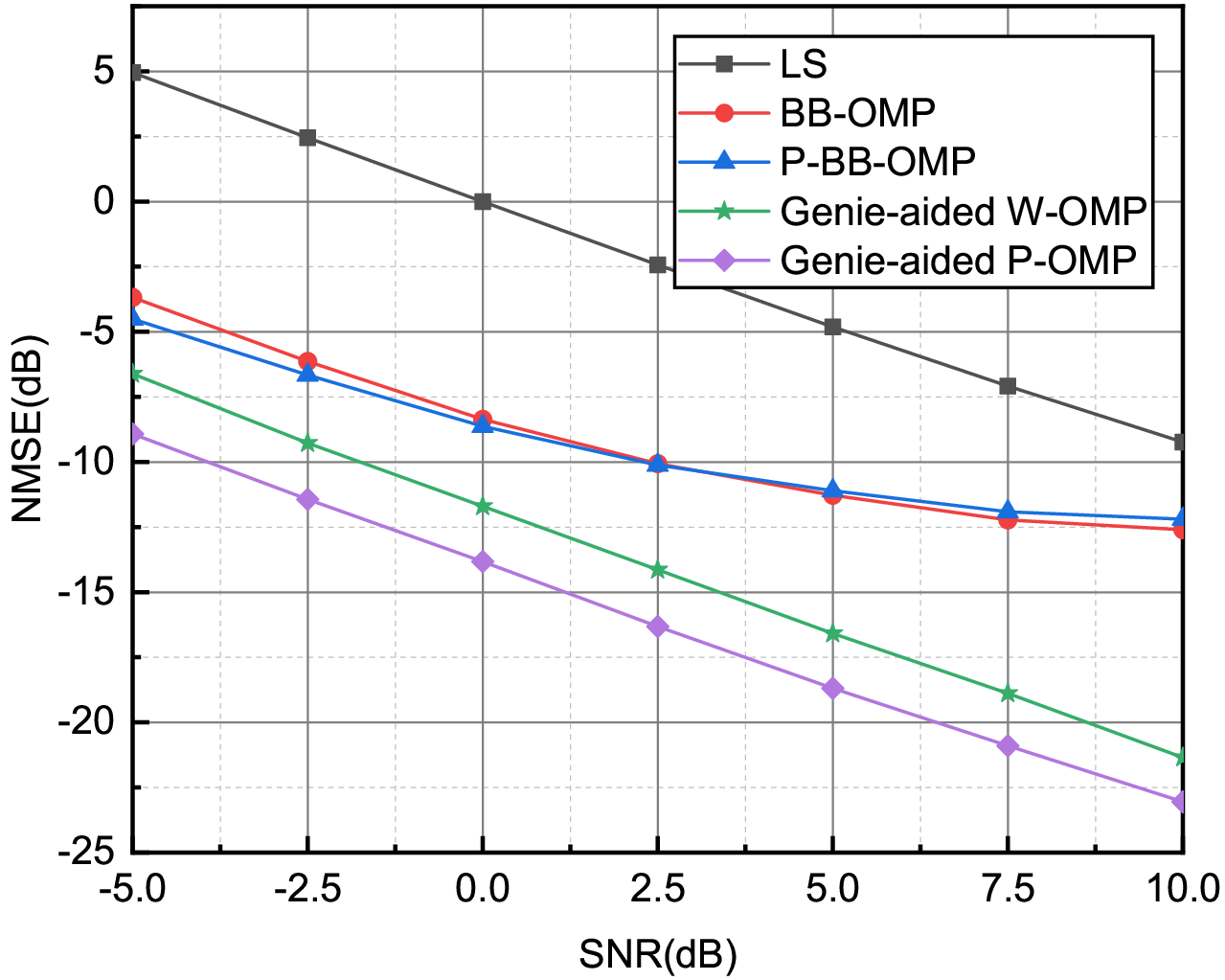}
		\label{NMSE4a}
	}
	\subfigure[$Q = 45$]{
		\includegraphics[width=0.22\textwidth]{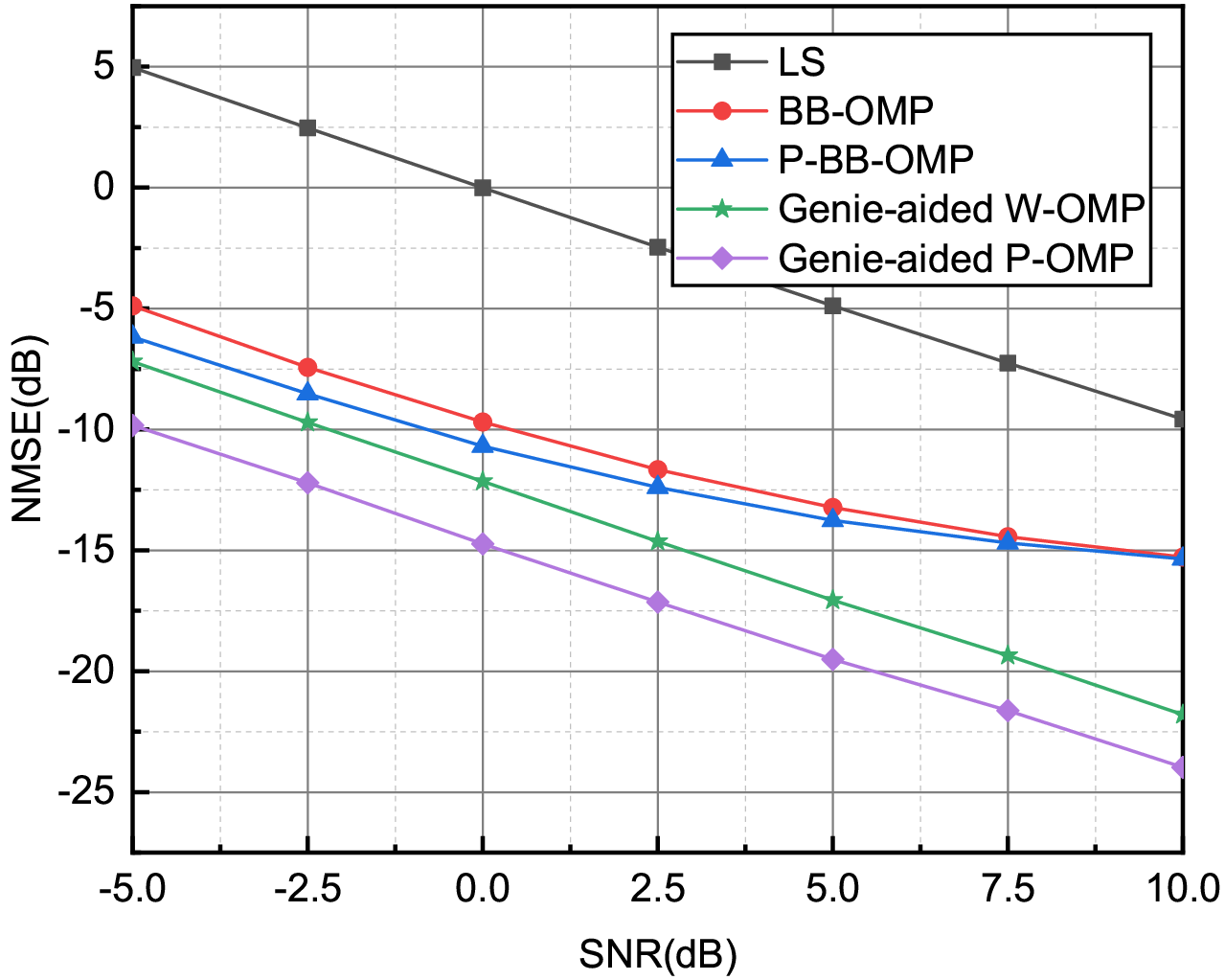}
		\label{NMSE4b}
	}
	\caption{NMSE versus different codebooks}
	\label{NMSE4}
\end{figure}

Inspired by the polar-main codebook proposed in~\cite{PolarCS}, we also try to utilize the polar-main codebook for replacing the wavenumber-domain codebook in the formula (\ref{Phi}) to evaluate the estimation performance. The estimation method based on the polar-domain codebook and belief of VR is called the P-BB-OMP algorithm.
Fig. \ref{NMSE4} illustrates the estimation performance comparison between the polar-domain and wavenumber-domain codebook with $\psi=0.25$. The results reveal that, under the assumption of perfect knowledge of VR, the polar-domain codebook achieves superior estimation performance. This superiority stems from the polar-domain ability to directly sample distance and angle domains, providing a more fine-grained spatial division, as opposed to the wavenumber-domain codebook, which only samples spatial frequency, with distance information implicit in the effective spatial bandwidth.
However, the performance advantage is constrained when VR information is not perfectly known, particularly with smaller pilot lengths. Additionally, it is noteworthy that the size of the polar-domain codebook is significantly larger than the wavenumber-domain codebook. For instance, when $N=256$ and $S=2$, the size of the polar-domain codebook is $2201$ \cite{PolarCS}, whereas the wavenumber-domain codebook is only $512$. Consequently, considering their reduced dimensions and adequate estimation performance, wavenumber-domain codebooks emerge as a practical choice for SnS XL-MIMO channels, particularly in LoS or sparse scattering environments.

\vspace{-1em}
\section{Conclusion and Future Works}
\label{section6}
In this paper, we have studied the CE problem in XL-MIMO systems with a hybrid precoding architecture, taking into account the spherical wavefront effect and the SnS property.
Initially, we assessed the impact of the SnS property on both the antenna domain and the wavenumber domain. Subsequently, we proposed a TS-VDCE scheme to fully exploit the spatial-domain and wavenumber-domain characteristics of SnS channels. Specifically, in the first stage, leveraging the spatial correlation of VR, 
the user's VR is effectively detected by the VRDO-MP algorithm. In the second stage, building upon the estimated VR information and exploiting wavenumber-domain sparsity, we devised the BB-OMP algorithm to estimate the SnS XL-MIMO channels. The simulation results demonstrated that the proposed TS-VDCE scheme can achieve significantly better VRER and NMSE performance than existing VR detection and CE schemes, with reduced pilot overhead, especially in low SNR scenarios. {Meanwhile, it is also applicable to different channel scenarios, including near-field spatial stationary scenario, near-field SnS scenarios, and far-field scenario.}
	
In this study, VR detection and CE are performed sequentially, potentially leading to error propagation. Therefore, the pursuit of concurrent VR detection and CE is an interesting avenue for further research. Furthermore, the SnS property is captured through a visibility indicator vector, whose dimension scales with the number of antenna elements. Consequently, with an increasing number of antennas, the parameters needed to describe SnS also grow significantly, leading to substantial estimation overhead. Hence, another key challenge is the efficient representation of SnS channels, aiming to use fewer parameters to accurately characterize the SnS property.
\bibliographystyle{IEEEtran}
\bibliography{ref}
\end{document}